 \documentstyle{article}

   \newcommand{\beqs}{\arraycolsep1.5pt\begin{eqnarray}}
   \newcommand{\eeqs}{\end{eqnarray}\arraycolsep5pt}
   \newcommand{\beqsn}{\arraycolsep1.5pt\begin{eqnarray*}}
   \newcommand{\eeqsn}{\end{eqnarray*}\arraycolsep5pt}
   \newcommand{\bmatrix}{\arraycolsep5pt\begin{array}}
   \newcommand{\ematrix}{\end{array}\arraycolsep1.5pt}

  \font\bfcal=cmbsy10

\newcommand{\bvarphi}{\mbox{\boldmath $\varphi$}}
\newcommand{\bpsi}{\mbox{\boldmath $\psi$}}

 \newcommand{\sgn}{\mathop{\rm sgn}\nolimits}
 \newcommand{\Res}{\mathop{\rm Res}\nolimits}
 \newcommand{\diag}{\mathop{\rm diag}\nolimits}
 \newcommand{\cut}{\llap{$-$}}

 \def\bbbone{{\mathchoice {\rm 1\mskip-4mu l} {\rm 1\mskip-4mu l}
 {\rm 1\mskip-4.5mu l} {\rm 1\mskip-5mu l}}}
 \def\bbbr{{\rm I\mskip -3.5mu R}}
 \def\bbbc{{\mathchoice {\setbox0=\hbox{$\displaystyle\rm C$}\hbox{\hbox
 to0pt{\kern0.4\wd0\vrule height0.9\ht0\hss}\box0}}
 {\setbox0=\hbox{$\textstyle\rm C$}\hbox{\hbox
 to0pt{\kern0.4\wd0\vrule height0.9\ht0\hss}\box0}}
 {\setbox0=\hbox{$\scriptstyle\rm C$}\hbox{\hbox
 to0pt{\kern0.4\wd0\vrule height0.9\ht0\hss}\box0}}
 {\setbox0=\hbox{$\scriptscriptstyle\rm C$}\hbox{\hbox
 to0pt{\kern0.4\wd0\vrule height0.9\ht0\hss}\box0}}}}
\begin{document}

%%%%%%%%%%%%%%%%%%%%%%%%%%%%%%%%%%%%%%%%%%%%%%%%%%%%%%%%%%%%%%%%%%%%%%%%%%%%%

\title{MULTIDIMENSIONAL LOCALIZED SOLITONS
\thanks{Work supported in part by Ministero delle Universit\`a e
della Ricerca Scientifica e Tecnologica, Italy}}
\author{M.~Boiti \and L.~Martina \and F.~Pempinelli\\
Dipartimento di Fisica dell'Universit\`a e
Sezione INFN\thanks{e-mail: BOITI@LECCE.INFN.IT, MARTINA@LECCE.INFN.IT and
PEMPI@LECCE.INFN.IT}\\ 73100 Lecce, ITALIA}
\date{August 25, 1993 (to be published in {\sl Chaos Solitons \& Fractals})}
\maketitle

\begin{abstract}
Recently it has been discovered that some nonlinear evolution equations in
$2+1$ dimensions, which are integrable by the use of the Spectral
Transform, admit localized (in the space) soliton solutions. This article
briefly reviews some of the main results obtained in the last five years
thanks to the renewed interest in soliton theory due to this discovery. The
theoretical tools needed to understand  the
unexpected richness of behaviour of multidimensional localized solitons
during their mutual scattering are furnished. Analogies and especially
discrepancies with the unidimensional case are stressed.
\end{abstract}

\section{Introduction}
\subsection{Special features of solitons in two
dimensions}

Since the discovery of the soliton in 1965 by Zabusky and Kruskal a large new
domain of mathematical physics developed and is believed to have reached
maturity. It is generically called the soliton theory. Its principal
mathematical tool is the so called Spectral (or Scattering) Transform that is
used to solve a large class of nonlinear evolution equations in 1+1 (one
spatial and one temporal) dimensions.

Some of these equations, in particular the nonlinear Schr\"odinger (NLS)
equation and its generalizations, can be obtained in an appropriate
multiscale limit from a very large class of nonlinear dispersive equations.
Therefore, it is not astonishing that applications of soliton theory
are percolating through the whole of physics, especially quantum field theory,
solid state physics, nonlinear optics, plasma physics and hydrodynamics, and
other natural sciences.

The most impressive phenomenon in the theory and in the applications is the
existence of solitons, i.e. (localized) coherent structures that
mutually interact preserving their individuality.

In the last decade many efforts have been made to extend the soliton theory to
nonlinear evolution equations in 2+1 (two spatial and one temporal)
dimensions.
In fact the Spectral Transform was extended to dispersive nonlinear evolution
equations  in 2+1 dimensions but it was generally admitted the lack of two
dimensional localized solitons. Only recently, in 1988, it has been discovered
(by Boiti, Leon, Martina and Pempinelli) that all the equations in the
hierarchy related to the Zakharov--Shabat hyperbolic spectral problem in the
plane have (exponentially) localized soliton solutions. The most
representative equation in the hierarchy is the Davey--Stewartson I equation
(DSI), which provides a two dimensional generalization of the NLS equation.

This discovery has stimulated a renewed interest in soliton theory. The first
results are very promising. In particular soliton solutions display a richer
phenomenology than in 1+1 dimensions. This opens the way to applications in
multidimensions, which, hopefully, are expected to be even more interesting
than in one space dimension.

In contrast with the 1+1 dimensional case the time evolution of the solution
of
the Davey--Stewartson equation is not uniquely determined by the initial data.
In addition one has to give, at all times, boundary data. If they are chosen
to
be identically zero solitons cannot be present. For a convenient
choice the solution can contain solitons but not necessarily does.
In the affirmative case the boundary data fix the
kinematics of the incoming and outgoing scattering solitons, i.e.
their velocities and locations in the plane in the far past and in the
far future. The initial data fix the dynamics of the interaction.

The scattering of the solitons can be inelastic and they can change
shape and also exchange mass (energy or charge according to the
specific physical interpretation). In fact, while the total mass of
solitons is conserved,  the mass of the single soliton, in general, is
not preserved by the interaction and solitons can also  simulate
inelastic scattering processes of quantum particles as creation  and
annihilation, fusion and fission, and interaction with virtual
particles.

\subsection{Guidelines for additional reading}

One main feature of the nonlinear dynamical systems is that
different non equivalent approaches are possible and each one is useful
and clarifying from a special point of view.

Also multidimensional
solitons can and, effectively, have been studied by using different
tools. In this article, mainly for lack of space, we made the
(questionable) choice of using only the B\"acklund transformations and
a special version of the Spectral Transform. In fact, we collected,
reorganized and simplified the results that the reader can find
scattered, with some minor additional details, in references
\cite{localized}--\nocite{localizedST,KPST,DSIST,%
bifurcation,bifurcationDubna90,wavesDSI}%
\cite{dynamics}.
In the first article of the list the localized solitons in the plane
have been discovered and in the second one a preliminary analysis of
the properties of the new Spectral Transform needed to describe them is
performed. These two papers opened the way to a deeper understanding of
integrable nonlinear evolution equations in multidimensions and to the
search and discovery of other dynamical systems admitting localized
coherent structures.

Another special version of the Spectral Transform theory different from
that one presented in this paper has been developed in
\cite{localizedFokasSantini}--%
\nocite{localizedSantini}\cite{localizedFokasSantinireview}.
This alternative theory is relevant because it has been used to clarify
the role played by the boundaries and to show that multidimensional
solitons, in contrast with the one dimensional solitons, can interact
inelastically. These authors suggested to call these solitons dromions in
order to stress the fact that they can be driven everywhere in the plane
along tracks (dromos in greek) by choosing a suitable motion of the
boundaries.

In both versions the Spectral Transform for
the nonstationary Schr\"odinger equation (with a potential vanishing at
large distances in every direction except a finite number) plays a
fundamental role. To extend the theory, originally developed for
potentials vanishing at large distances \cite{FokasAblowitzKPI}, to
this case it has been necessary to introduce a new mathematical entity
resembling the resolvent of the linear operator theory. For lack of
space this topic has been skipped in this article.  The interested
reader can consult \cite{KPline,KPrussian}. The multi--soliton solution
of the Kadomtsev--Petviashvili I (KPI) equation, that is needed in order to
build the multi--soliton solution of the DSI and DSIII systems, has been
derived in \cite{KPsolitonsfirst} and in its most general form in
\cite{generalKPsolitons}.

Also other
methods have been used to get the multidimensional solitons, the
bilinear approach \cite{localizedHietarinta}, the quantum machinery of
creation and annihilation operators
\cite{localizedAlonsoJaulent,localizedAlonso}, the Grammian
determinants \cite{localizedNimmo}, the Darboux transformations
\cite{localizedNimmo2} and the dressing method \cite{FZdressing}.  The most
general form of the
multi--soliton solution together with the remark that the number of
solitons is not necessarily conserved were first given in
\cite{localizedHietarinta}.

The search of multidimensional solitons has been extended by different
authors to other nonlinear evolution equations in $2+1$
dimensions
\cite{localizedKonopel:a}--\nocite{localizedKonopel:b,localizedNizhnikNV}%
\cite{localizedSchief}.
Some interesting attempts using the $\overline\partial$--method and a
direct method have been done also in higher dimensions, precisely in
\cite{Leon3+1} and
\cite{SabatierN+1:a}--\nocite{SabatierN+1:b}\cite{DegasperisN+1}.

In
\cite{ZakharovKuznetsov}--%
\nocite{Calogeromultiscale,CalogeromultiscaleI,CalogeromultiscaleII}%
\cite{CalogeroMaccari} it has been shown that the DSI and, successively in
\cite{CalogeroExeter}, that also the DSIII system can be obtained by
using a multiscale limit starting from a very large class of dispersive
nonlinear equations. In particular the DSI equation with boundaries
compatible with solitons can be embedded into the KPI equation
\cite{DSfromKP}. These results open the way to the search of physical
applications other than hydrodynamics, studied in \cite[and references
quoted therein]{AblowitzSegurfluid,SautDS}.  In this respect a more
discouraging analysis is made in \cite{KaupDSI}.  This last paper,
mainly dedicated to the consequences of the lack of conserved
quantities for DSI, contains also an interesting examination of the
literature dealing with the discovery of the DSI system and with the
early attempts to build the corresponding Spectral Transform.

For the
Hamiltonian version of the DSI system and its quantum extension see
\cite{SchultzDSH:a}--\nocite{SchultzDSH:b,VillarroelDSH,KulishDSH:a}%
\cite{KulishDSH:b}. A Hamiltonian version of the DSIII system appears
already in
\cite{ShulmanDSIII} and a bi--Hamiltonian version in
\cite{SantiniFokasDSIII}. For the general method for building integrable
nonlinear evolution equations starting from a $\overline\partial$--problem
that allowed to prove that DSIII is integrable see
\cite{Sabatiermethod:a}--\nocite{Sabatiermethod:b}\cite{Sabatiermethod:c}
and, specifically, \cite{DSIII}.  For more details on the localized soliton
solutions of DSIII see \cite{DSIIIExeter92}. For the DSII system which is
related to the elliptic version of the Zakharov--Shabat problem in the plane
see \cite{FokasDS,FokasAblowitzDS} for the Spectral Transform theory and
\cite{PogrebkovKPII:a,PogrebkovKPII:b} for the singular soliton solutions.

Finally, between the different excellent existing books on solitons let
us suggest to the reader \cite{Calogerobook} and \cite{Faddeevbook}, as
the most accurate and complete for the nonlinear evolution equations
associated to the Schr\"odinger and to the Zakharov--Shabat
spectral equation in 1+1 dimensions, respectively, and
\cite{Clarksonbook} as the most updated and comprehensive. In
particular for those who want to have a general overlooking on the
subject and a rich bibliography to pick over this book is particularly
recommended. A new book on multidimensional soliton has been just
announced \cite{Konopelbook:b}.  For the B\"acklund and Darboux
transformations see \cite{Rogersbook} and \cite{Matveevbook}. Special
attention to the algebraic and geometric approach to the soliton
equations has been dedicated in \cite{Konopelbook:a} and to the
multidimensional case in \cite{Konopelbook:a2}.

In all these books the reader can
find many useful references for the problems considered in this paper
and for related problems.  Here we want to quote only some
references more relevant historically or more close to our specific
approach.  Precisely \cite{ZabuskyKruskal:a,ZabuskyKruskal:b} for the
discovery of the one dimensional solitons, \cite{ZS,AKNS} for the extension
of the theory to the nonlinear Schr\"odinger equation,
\cite{BealsCoifmanDBAR:a}--\nocite{BealsCoifmanDBAR:b}%
\cite{BealsCoifmanreview} for the introduction of the so--called
$\overline\partial$--method,
\cite{ZakharovManakov2+1:a}--\nocite{ZakharovManakov2+1:b}%
\cite{Manakov2+1}, \cite{FokasAblowitzDS},
\cite{FokasAblowitzKPI},
\cite{AblowitzYaakovFokas}--%
\nocite{2+1KdV,2+1Schrodinger,Shine-Gordon,2+1longwaves}%
\cite{FokasSung} for the extension of the Spectral Transform to
$2+1$ dimensions, \cite{2+1KdV}--\cite{2+1longwaves} in particular for the
introduction of the ``weak'' Lax representation of the integrable equations
in $2+1$ dimensions, \cite{CalogeroBT:a,CalogeroBT:b} for the B\"acklund
transformations and \cite{2+1BT} for the extension of the B\"acklund
transformations to $2+1$ dimensions.

By offering these guidelines for recovering interesting references we
hope to redress the omissions due to oversight that have certainly
occurred. Anyway, we apologize for this to the reader as well to the
authors who may have unjustifiably excluded.

\subsection{The Davey--Stewartson I equation}

The Davey--Stewartson (DS) systems model the evolution of weakly nonlinear
water
waves that travel predominantly in one direction, are nearly monochromatic and
are slowly modulated in the two horizontal directions. We are interested in
the
special DS system (DSI equation) that one gets in the shallow water limit when
the effects of the surface tension are important. In characteristic
coordinates
and dimensionless form the DSI equation is a system of two coupled equations
\beqs
\label{physicalDSI}
iq_t+q_{uu}+q_{vv}-(\varphi_u+\varphi_v-\sigma_0\vert q\vert^2)q=0\\
2\varphi_{uv}=\sigma_0(\vert q\vert^2)_u+\sigma_0(\vert q\vert^2)_v,
\quad\sigma_0=\pm 1\nonumber
\eeqs
where $q(u,v,t)$ is the (complex) envelope of the free surface of the water
wave we are considering and $\varphi(u,v,t)$ is the (real) velocity potential
of the mean motion generated by the surface wave.

It is worth to stress that the DSI system is not necessarily placed in the
context of water waves. Indeed, it has been shown that a very large class of
nonlinear dispersive equations in 2+1 dimensions reduces in an appropriate
asymptotic limit to the DSI equation and therefore we expect it to arise in
many different physical applications.

To exhibit explicitly the boundary value of $\varphi$ at large
distance in the $(u,v)$ plane allowed by the second equation in
(\ref{physicalDSI}) it is convenient to introduce the two fields
\beqs
A^{(1)}&=&-\varphi_v+{1\over 2}\sigma_0\vert q\vert^2\\
A^{(2)}&=&\varphi_u-{1\over 2}\sigma_0\vert q\vert^2\nonumber
\eeqs
and to rewrite the DSI equation as
\begin{equation}
\label{reducedDSI}
iq_t+q_{uu}+q_{vv}+(A^{(1)}-A^{(2)})q=0
\end{equation}
\beqsn
A^{(1)}&=&-{1\over 2}\int_{-\infty}^u\!\!du'\,
\sigma_0(\vert q\vert^2)_v+a_0^{(1)}(v,t) \\
A^{(2)}&=&{1\over 2}\int_{-\infty}^v\!\!dv'\,
\sigma_0(\vert q\vert^2)_u+a_0^{(2)}(u,t)
\eeqsn
where $a_0^{(1)}$ and $a_0^{(2)}$) are the arbitrary boundaries.

In fact we could make another not equivalent choice and write the DSI equation
as
\begin{equation}
\label{HamiltonDSI}
iq_t+q_{uu}+q_{vv}+(A^{(1)}-A^{(2)})q=0
\end{equation}
\beqsn
A^{(1)}&=&-{1\over 4}\left(\int_{-\infty}^u+\int_{+\infty}^u\right)du'\,
\sigma_0(\vert q\vert^2)_v+A_0^{(1)}(v,t) \\
A^{(2)}&=&{1\over 4}\left(\int_{-\infty}^v+\int_{+\infty}^v\right)dv'\,
\sigma_0(\vert q\vert^2)_u+A_0^{(2)}(u,t)
\eeqsn
where now $A_0^{(1)}(v,t)$ and
$A_0^{(2)}(u,t)$ are the arbitrary boundaries.

It can be shown that the proper boundary conditions to be chosen are dictated
by the specific multiscale limit one is choosing in getting the DSI equation.
Specifically, equation (\ref{reducedDSI}) can be obtained via a multiscale
limit from the Kadomtsev--Petviashvili (KPI) evolution equation while
maintaining
well posedness in time.

In order to solve the DSI equation it is convenient to introduce its more
general two component version
\begin{equation}
\label{QDSI}
iQ_t+\sigma_3(Q_{uu}+Q_{vv})+[A,Q]=0
\end{equation}
where
\begin{equation}
Q= \left(\matrix
{ 0        & q(u,v,t) \cr
  r(u,v,t) &     0    \cr  } \right)
\end{equation}
and
\begin{equation}
A= \left(\matrix
{ A^{(1)}  &     0    \cr
     0     &  A^{(2)} \cr  } \right)
\end{equation}
with
\beqs
\label{ADSI}
A^{(1)}&=&-{1\over2}\int_{-\infty}^udu'\,(Q^2)_v+a_0^{(1)}(v,t)\\
A^{(2)}&=&{1\over2}\int_{-\infty}^vdv'\,(Q^2)_u+a_0^{(2)}(u,t)\nonumber
\eeqs
or
\beqs
\label{HamiltonADSI}
A^{(1)}=-{1\over4}(\int_{-\infty}^u+\int_{+\infty}^u)du'\,
(Q^2)_v+A_0^{(1)}(v,t)\\
A^{(2)}={1\over4}\left(\int_{-\infty}^v+\int_{+\infty}^v\right) dv'\,
(Q^2)_u+A_0^{(2)}(u,t)\nonumber
\eeqs
according to the boundary conditions we choose. The previous considered
equations are simply obtained for
\begin{equation}
\label{reduction}
r=\sigma_0\bar{q}
\end{equation}
(where $\bar{q}$ denotes the complex conjugate of $q$) and are called reduced
DSI equations in contrast with (\ref{QDSI}), (\ref{ADSI}) and
(\ref{HamiltonADSI}) which can be named DSI equations with no additional
specification.

The DSI equation can be obtained as the compatibility condition (Lax
representation)
\begin{equation}
\label{1.11}
[T_1,T_2]=0
\end{equation}
for two underlying linear operators
\beqs
\label{1.12}
T_1(Q)&=&\partial_x+\sigma_3\partial_y+Q\\
T_2(Q)&=&i\partial_t+\sigma_3\partial_y^2+Q\partial_y-{1\over2}\sigma_3Q_x+
{1\over2}Q_y+A
\label{1.13}
\eeqs
where
\begin{equation}
x={1\over2}(u+v),\qquad y={1\over2}(u-v).
\end{equation}
and $\sigma_3$ is the Pauli matrix.

The first of these operators is the Zakharov--Shabat hyperbolic spectral
operator in the plane and can be considered to define a linear spectral
problem
in which $Q$ plays the role of the data. In the case (\ref{ADSI}) with
$a_0^{(1)}$ and $a_0^{(2)}$ considered as boundaries given at all times one
can
define the Spectral Transform of $Q$ and solve the initial value problem for
the DSI. This is explicitly done in section 3.

In the case (\ref{HamiltonADSI}) one can introduce the Hamiltonian
\begin{equation}
H=\!\!\int\!\!\!\int\!\!dudv\left[r(\partial_u^2+\partial_v^2)q-{1\over4}qr
(\partial_u\partial_v^{-1}+
\partial_v\partial_u^{-1})qr+(A_0^{(1)}-A_0^{(2)})qr\right]
\end{equation}
and the canonical Poisson brackets
\begin{equation}
\{F,G\}=i\!\!\int\!\!\!\int\!\!dudv\left[
{\delta F\over\delta q}{\delta G\over\delta r}-
{\delta F\over\delta r}{\delta G\over\delta q}\right]
\end{equation}
where $q$ and $r$ are the conjugate variables. Then the equations of motion
\begin{equation}
q_t=\{q,H\},\qquad r_t=\{r,H\}
\end{equation}
yield the DSI equation. The problem of defining a Spectral Transform is
completely open and, moreover, only in the special case $A_0^{(1)}\equiv
A_0^{(2)}\equiv 0$ it has been shown that the DSI equation is Hamiltonian with
a continuous infinity of independent commuting constants and is completely
integrable in the Hamiltonian sense. This case is usually named Hamiltonian
case.

\subsection{The Davey--Stewartson III equation}
There is another nonlinear evolution equation that can be associated to
the Zakharov--Shabat hyperbolic spectral operator in the plane and that
admits localized soliton solutions.

To get it we need to introduce a weaker form of the Lax representation
(\ref{1.11}). Precisely, we search for a second spectral operator $T_2$
that commute with the Zakharov--Shabat spectral operator $T_1$ in
(\ref{1.12}) only on the subspace of the eigenfunctions of $T_1$
(``weak'' Lax representation)
\begin{equation}
T_1\psi=0,\qquad [T_1,T_2]\psi=0.
\end{equation}
For
\beqs
&&T_1=2\diag(\partial_u,\partial_v)+Q\\
\label{T2weakDSIII}
&&T_2=i\partial_t+\partial_u^2+\partial_v^2+A+
\left(\bmatrix{cc}0&q_u\\r_v&0\ematrix\right)
\eeqs
we have the so--called DSIII system
\beqs
&&iQ_t+\sigma_3\left(Q_{vv}-Q_{uu}\right)+[A,Q]=0\\
&&A^{(1)}_u=-{1\over2}{\left(Q^2\right)}_{v}\\
&&A^{(2)}_v=-{1\over2}{\left(Q^2\right)}_{u}
\eeqs
to be compared with the DSI system in (\ref{QDSI}) and (\ref{ADSI}) or
(\ref{HamiltonADSI}). The DSIII system. as DSI, is compatible with the
reduction (\ref{reduction}).

Of course also DSI can be obtained by using a ``weak'' Lax
representation instead of the usual ``strong'' one in (\ref{1.11}).
Then, one can choose for $T_2$ instead of (\ref{1.13})
\begin{equation}
\label{T2weakDSI}
T_2=i\partial_t+\partial_v^2-\partial_u^2+A+
\left(\bmatrix{cc}0&-q_u\\r_v&0\ematrix\right)
\end{equation}
and get again DSI.
The theory of the B\"acklund transformation and of the Spectral
Transform developed in the following sections does not change if the
starting Lax pair is ``weak'' or ``strong''.  Therefore, comparison of
definition (\ref{T2weakDSIII}) with (\ref{T2weakDSI}) makes
clear that one cannot expect any difference, apart some signs, between
formulae for DSI and DSIII.

In particular the time evolution of $\psi$ for DSIII has to be fixed as
follows
\begin{equation}
T_2\psi=-k^2\psi
\end{equation}
to be compared with (\ref{4.10}) for DSI.

The DSIII system admits for convenient boundaries
of the form
\beqs
A^{(1)}&=&-{1\over2}\int_{-\infty}^udu'\,(Q^2)_v+a_0^{(1)}(v,t)\\
A^{(2)}&=&-{1\over2}\int_{-\infty}^vdv'\,(Q^2)_u+a_0^{(2)}(u,t)\nonumber
\eeqs
localized solitons of the same shape of those of DSI, which exhibit similar
dynamical phenomena but evolve differently in time.

We write here the one soliton solution
\begin{equation}
q=-\frac{2\lambda_\Im\eta\exp[i\varphi]}{D},\qquad
r=-\frac{2\mu_\Im\rho\exp[-i\varphi]}{D}
\end{equation}
where
\beqs
&&\varphi=\mu_\Re u+\lambda_\Re v+(\lambda_\Im ^2-
\lambda_\Re ^2-\mu_\Im ^2+\mu_\Re ^2)t,\\
&&D=2\gamma(\cosh\xi_1+\cosh\xi_2)+\exp(\xi_2),
\qquad\gamma=\hbox{$1\over4$}\eta\rho,\\
&&\xi_1=-\mu_\Im
u-\lambda_\Im v+2(\lambda_\Im\lambda_\Re-\mu_\Im\mu_\Re)t,\\
&&\xi_2=\mu_\Im
u-\lambda_\Im v+2(\lambda_\Im\lambda_\Re+\mu_\Im\mu_\Re)t
\eeqs
to be compared with (\ref{3.36})--(\ref{3.38.b}). As in the DSI case
the complex parameters  $\lambda=\lambda_\Re+i\lambda_\Im$ and
$\mu=\mu_\Re+i\mu_\Im$ are the discrete eigenvalues of the
associated Zakharov--Shabat spectral problem and $\rho$ and
$\eta$ are arbitrary complex constants satisfying the conditions
$\gamma\in\bbbr$ and $\gamma(1+\gamma)>0$. The multi--soliton solution can
be easily obtained by taking the same lines we choose in the following
sections for the DSI equation.

\section{Solitons via B\"acklund Transformations}
\setcounter{equation}{0}

The B\"acklund transformations have their origin in work by B\"acklund
in
the late nineteenth century and are, therefore, the oldest tool used in
exploring nonlinear integrable systems. Much more recently many different
sophisticated and powerful methods have been developed, in particular the
Spectral Transform and the dressing method. However, in our opinion, the
B\"acklund transformations remain the simplest way for getting the soliton
solutions. Moreover, because, under appropriate circumstances, a reiterated
application of the B\"acklund transformations generate a sequence of solutions
by a purely algebraic superposition principle they can be used to study the
interaction properties of the solitons.

The simplest way to generate a B\"acklund transformation is to use the gauge
invariance of the linear spectral problem associated to the nonlinear
evolution
equation one is considering. The gauge that generates the B\"acklund
transformation is called B\"acklund gauge. The localized soliton solutions of
the DSI equation, with boundaries of the form in (\ref{ADSI}), were for the
first time derived by using these special gauge transformations. Successively
they have been rederived by using the techniques of the Spectral Transform.
But, in the case of the Hamiltonian DSI equation, we have not, presently, at
our disposal the Spectral Transform or the dressing method and, consequently,
in order to get explicit solutions we are left with the necessity to
generalize
the B\"acklund gauges such as to include also the special form of the
boundaries in (\ref{HamiltonADSI}). To choosing these boundaries corresponds
to imposing to the solutions nonlinear constraints, which can be solved only
by using the additional freedom at our disposal in the generalized
B\"acklund gauges. We are able to write explicitly infinite wave solutions
with constant and periodically modulated amplitudes.

\subsection{Generalized B\"acklund gauge transformation}

Once given a solution $Q$ of the DSI equation we want to generate a new
solution
$Q'$ of the same equation by using a convenient gauge operator $B$ that
transforms according to the equation
\begin{equation}
\label{gaugeB}
\psi'=B(Q',Q)\psi
\end{equation}
the matrix solution $\psi$ of the principal spectral equation
\begin{equation}
T_1(Q)\psi=0
\end{equation}
for $Q$ to the matrix solution $\psi'$ of the same spectral problem
\begin{equation}
T_1(Q')\psi'=0
\end{equation}
for $Q'$.

It is easy to verify that if $B$ satisfies
\beqs
\label{spaceBacklund}
T_1(Q')B(Q',Q)-B(Q',Q)T_1(Q)=0\\
\label{timeBacklund}
T_2(Q')B(Q',Q)-B(Q',Q)T_2(Q)=0
\eeqs
then $T_1(Q')$ and $T_2(Q')$ satisfy the same compatibility condition
\begin{equation}
[T_1,T_2]=0
\end{equation}
as $T_1(Q)$ and $T_2(Q)$ and, consequently, $Q'$ satisfies the DSI equation.
In
this case the gauge $B$ is called B\"acklund gauge and the equations
(\ref{spaceBacklund}) and (\ref{timeBacklund}) yield, respectively, the
so--called space and time component of the B\"acklund transformation.

Non trivial B\"acklund gauges are polynomial in the operator $\partial_y$. We
are interested in the most general B\"acklund gauge of first order of the form
\begin{equation}
B(Q',Q)=\alpha\partial_y+B_0(Q',Q)
\end{equation}
with $\alpha$ a constant diagonal matrix and $B_0$ a matrix. By inserting it
in
(\ref{spaceBacklund}) we get its functional form
\begin{equation}
\label{fullBacklund}
B(Q',Q)=\alpha\partial_y-\hbox{$1\over2$}\sigma_3(Q'\alpha-\alpha Q)
-\hbox{$1\over2$}\sigma_3\alpha\, {\cal I}(Q'^2-Q^2)+\beta
\end{equation}
and the space component of the B\"acklund transformation
\beqs
Q'\big[\beta-\hbox{$1\over2$}\alpha\sigma_3\,{\cal I}(Q'^2-Q^2)\big]
-\big[\beta-\hbox{$1\over2$}\alpha\sigma_3\,{\cal
I}(Q'^2-Q^2)\big]Q\nonumber\\
{ }-\hbox{$1\over2$}\sigma_3(Q'\alpha-\alpha Q)_x
-\hbox{$1\over2$}(Q'\alpha+\alpha Q)_y=0.\label{spacecomponent}
\eeqs
The matrix operator ${\cal I}$ is defined by
\begin{equation}
{\cal I}=(\partial_x+\sigma_3\partial_y)^{-1}
\end{equation}
and the diagonal matrix $\beta$ is subjected to the constraint
\begin{equation}
\label{beta}
(\partial_x+\sigma_3\partial_y)\beta=0,
\end{equation}
i.e. it is of the form
\begin{equation}
\beta=
\left(\matrix{\beta_1(v,t)  &       0        \cr
                    0         & \beta_2(u,t) \cr}  \right)
\end{equation}
where
\begin{equation}
u=x+y,\qquad v=x-y
\end{equation}
and $\beta_1$ and $\beta_2$ are arbitrary functions. Note that, in contrast
with the 1+1 dimensional case, the matrix $\beta$ that plays the role of
`constant of integration' in the solution of (\ref{spaceBacklund}) admits also
a space dependence. This additional freedom will be used in the following for
getting soliton solutions of the Hamiltonian DSI equation.

By inserting $B(Q',Q)$ in (\ref{timeBacklund}) we get the time component of
the
B\"acklund transformation, which can be shown by use of (\ref{spacecomponent})
to be equivalent to the DSI for $Q'$, and two additional partial differential
equations
\begin{equation}
\label{timecomponentBT}
{\big[\beta-\hbox{$1\over2$}\alpha\sigma_3\,{\cal I}(Q'^2-Q^2)\big]}_y
+\hbox{$1\over2$}\alpha\sigma_3(A'-A)+\hbox{$1\over4$}\alpha(Q'^2-Q^2)=0
\end{equation}
\beqs
i{\big[\beta-\hbox{$1\over2$}\alpha\sigma_3\,{\cal I}(Q'^2-Q^2)\big]}_t+
(A'-A)\big[\beta-\hbox{$1\over2$}\alpha\sigma_3\,{\cal
I}(Q'^2-Q^2)\big]\nonumber\\
-\hbox{$1\over2$}\alpha{(A'+A)}_y+\hbox{$1\over4$}\alpha
\sigma_3{(Q'^2+Q^2)}_y
-\hbox{$1\over4$}\big[(\partial_x+\sigma_3\partial_y)Q'\big]Q'
\alpha\nonumber\\
-\hbox{$1\over4$}\alpha
Q\big[(\partial_x-\sigma_3\partial_y)Q\big]+\hbox{$1\over4$}
(\partial_x-\sigma_3
\partial_y)
(Q'\alpha Q)=0.
\eeqs
These two equations can be used for determining the field $A'$ and the
admissible $\beta$'s. It can be verified that they are compatible with the
equation (\ref{beta}) for $\beta$ by applying the operator
$(\partial_x+\sigma_3\partial_y)$ to both of them and by showing that the two
obtained equations are identically satisfied for $Q$ and $Q'$ solutions of
the DSI equation and of the
space component of the B\"acklund transformation.

The B\"acklund gauge $B(Q',Q)$ for general $\alpha$ and $\beta$ in
(\ref{fullBacklund}) can be obtained by composing two simpler B\"acklund
gauges
that are called elementary B\"acklund gauges of the first and second kind.
They are obtained by choosing, respectively,
\begin{equation}
\label{alphabetaI}
\alpha_I=
\left(\matrix{1  &  0 \cr
              0  &  0 \cr}
\right),
\qquad \beta_I=
\left(\matrix{\lambda(v,t)  &  0 \cr
                   0        &  1 \cr}
\right)
\end{equation}
and
\begin{equation}
\label{alphabetaII}
\alpha_{II}=
\left(\matrix{0  &  0 \cr
              0  &  1 \cr}
\right),
\qquad \beta_{II}=
\left(\matrix{     1  &       0 \cr
                   0  &  \mu(u,t) \cr}
\right),
\end{equation}
and are noted $B_{I}(Q',Q;\lambda)$ and $B_{II}(Q',Q;\mu)$. They are not
compatible with the reduction
\begin{equation}
r=\sigma_0\bar q
\end{equation}
but by composing two elementary B\"acklund gauges of different kind one can
get
finally a solution satisfying the reduction.

This procedure simplifies radically the computation because, according to the
general feature of the B\"acklund transformations, the recursive application
of
the B\"acklund transformations can be achieved by purely algebraic means. An
additional simplification is obtained by imposing the commutativity of the
diagram
\begin{center}
\setlength{\unitlength}{0.03cm}
\begin{picture}(100,100)
\put(10,60){\vector(1,1){30}}
\put(60,90){\vector(1,-1){30}}
\put(10,40){\vector(1,-1){30}}
\put(60,10){\vector(1,1){30}}
\put(0,48){\shortstack{$Q$}}
\put(45,100){\shortstack{$Q_I$}}
\put(45,0){\shortstack{$Q_{II}$}}
\put(95,48){\shortstack{$Q'$}}
\put(-7,75){\shortstack{$B_I(\lambda)$}}
\put(80,75){\shortstack{$B_{II}(\mu)$}}
\put(-7,15){\shortstack{$B_{II}(\mu)$}}
\put(80,15){\shortstack{$B_I(\lambda)$}}
\end{picture}
\end{center}
which represents symbolically the following equation
\beqs
\label{3.17}
B(Q',Q;\lambda,\mu)&=&B_{II}(Q',Q_I;\mu)B_I(Q_I,Q;\lambda)\nonumber\\
&=&B_I(Q',Q_{II};\lambda)B_{II}(Q_{II},Q;\mu)
\eeqs
where $B(Q',Q;\lambda,\mu)$ is the B\"acklund gauge in (\ref{fullBacklund})
with
\begin{equation}
\alpha={\bbbone },
\qquad \beta=
\left(\matrix{\lambda(v,t)  &  0 \cr
                   0        &  \mu(u,t) \cr}
\right).
\end{equation}

\subsection{Localized soliton solutions}

The most natural choice for the parameters entering (\ref{alphabetaI})
and (\ref{alphabetaII})
is of course to take $\lambda$ and $\mu$ constants, say
\begin{equation}
\lambda(v,t)=i\lambda,\qquad \mu(u,t)=i\mu,
\label{3.18}
\end{equation}
where $\lambda$ and $\mu$ are complex constants. Then
we choose the starting solution $Q$ of (\ref{QDSI}) to be zero together
with its related auxiliary field $A$ (note, however, that the vanishing
of $Q$ does not imply the vanishing of $A$).

With these choices the commutation relation (\ref{3.17}) becomes
\begin{equation}
\label{BQOlambdamu}
B(Q,0;\lambda,\mu)=B_{II}(Q,Q_I;\mu)B_I(Q_I,0;\lambda)=
B_I(Q,Q_{II};\lambda)B_{II}(Q_{II},0;\mu).
\label{3.19}\end{equation}

By equating to zero the coefficients of the powers of $\partial_y$ it results
that
\begin{equation}
Q_I=
\left(\matrix{0  &  0 \cr
             r_I &  0 \cr} \right),
\qquad Q_{II}=
\left(\matrix{     0  &  q_{II} \cr
                   0  &     0     \cr}  \right),
\qquad Q=
\left(\matrix{     0  &  q  \cr
                   r  &  0     \cr}  \right),           \label{3.20}
\end{equation}
where
\begin{equation}
r_I=\rho(u,t)\exp(-i\lambda v),\qquad
q_{II}=\eta(v,t)\exp(i\mu u),                          \label{3.21}
\end{equation}
\begin{equation}
q={-q_{II,v}+i\lambda q_{II}\over
1+\hbox{$1\over4$}r_Iq_{II}},\qquad
r={r_{I,u}+i\mu r_I\over
1+\hbox{$1\over4$}r_Iq_{II}}
\label{3.22}
\end{equation}
and that the operator $\cal I$ satisfies the equation
\begin{equation}
\label{IQ2}
{\cal I}(Q^2)={1\over2}Q(Q_{I}-Q_{II}).
\end{equation}
Here above $\rho$ and $\eta$ are arbitrary functions of the space
variable, that can be represented by the following transforms
\begin{equation}
\rho(u,t)=\int\!\!\int dk\wedge d\bar{k}\;e^{-iku}\tilde\rho(k,t),
                                                        \label{3.23}
\end{equation}
\begin{equation}
\eta(v,t)=\int\!\!\int dk\wedge d\bar{k}\;e^{ikv}\tilde\eta(k,t),
                                                        \label{3.24}
\end{equation}
with the explicit time dependencies
\begin{equation}
\tilde\rho(k,t)=\tilde\rho(k,0)\exp[i(k^2+\lambda^2)t],\label{3.25}
\end{equation}
\begin{equation}
\tilde\eta(k,t)=\tilde\eta(k,0)\exp[-i(k^2+\mu^2)t],  \label{3.26}
\end{equation}
where $\tilde\rho(k,0)$ and $\tilde\eta(k,0)$ are arbitrary.

Everything now lies in the choice of these two arbitrary functions. We
prove hereafter that the requirement that the solution $Q$ is
localized and obeys the reduction condition (\ref{reduction})
determines uniquely the structure of the arbitrary functions $\tilde\rho$
and $\tilde\eta$. To that end we redefine
\begin{equation}
\rho(u,0)=2e^{-i\mu u}S(\sigma),                       \label{3.27}
\end{equation}
\begin{equation}
\eta(v,0)=2e^{i\lambda v}T(\tau),                    \label{3.28}
\end{equation}
with the new variables
\begin{equation}
\sigma={-i\over\mu-\bar{\mu}}\exp[i(\mu-\bar{\mu})u], \label{3.29}
\end{equation}
\begin{equation}
\tau={-i\sigma_0\over\lambda-\bar{\lambda}}
\exp[-i(\lambda-\bar{\lambda})v].                        \label{3.30}
\end{equation}

In these variables, the reduction condition (\ref{reduction}) takes the
following
form
\begin{equation}
[1+\bar S(\sigma)\bar T(\tau)]{dS\over d\sigma}=
[1+S(\sigma)T(\tau)]{d\bar T\over d\tau},               \label{3.31}
\end{equation}
which can be solved explicitly and admits 4 independent solutions. The
requirement of the localization selects the only solution
\beqs\bmatrix{c}
S=a\sigma+b,\qquad b\bar a-\bar ba=0,\\
T=\bar a\tau+c,\qquad ca-\bar c\bar a=0,
\ematrix                                             \label{3.32}\eeqs
where $a,b$ and $c$ are complex constants. Then, by means of (\ref{3.27}) and
(\ref{3.28}), we arrive at
\begin{equation}
\tilde\rho(k,0)=\rho[\delta(k-\mu)+\delta(k-\bar\mu)],\label{3.33}
\end{equation}
\begin{equation}
\tilde\eta(k,0)=\eta[\delta(k-\lambda)+\delta(k-\bar\lambda)].
                                                        \label{3.34}
\end{equation}

Here above $\eta$ and $\rho$ are arbitrary constants. The requirement of
the reduction condition gives the following constraint on these
constants
\begin{equation}
\rho(\mu-\bar\mu)=\sigma_0\bar\eta(\lambda-\bar\lambda).
                 \label{3.35}
\end{equation}

Therefore the localized one-soliton solution to the system (\ref{QDSI})
(\ref{ADSI}) can be written
\begin{equation}
q=-\frac{2\lambda_\Im\eta\exp[i\varphi]}{D},\qquad
r=-\frac{2\mu_\Im\rho\exp[-i\varphi]}{D},                \label{3.36}
\end{equation}
with the following definitions
\begin{equation}
\varphi=\mu_\Re u+\lambda_\Re v+(\lambda_\Im ^2-
\lambda_\Re ^2+\mu_\Im ^2-\mu_\Re ^2)t,
                                                        \label{3.37}
\end{equation}
\begin{equation}
D=2\gamma(\cosh\xi_1+\cosh\xi_2)+\exp(\xi_2),
\qquad\gamma=\hbox{$1\over4$}\eta\rho,
\label{3.38.a}
\end{equation}
\begin{equation}
\xi_1=-\mu_\Im
u-\lambda_\Im v+2(\lambda_\Im\lambda_\Re+\mu_\Im\mu_\Re)t,
\end{equation}
\begin{equation}
\xi_2=\mu_\Im
u-\lambda_\Im v+2(\lambda_\Im\lambda_\Re-\mu_\Im\mu_\Re)t,
                                                        \label{3.38.b}
\end{equation}
and where $\lambda_\Re$ denotes the real part of $\lambda$ and $\lambda_\Im$
its imaginary part.

For $\lambda_\Im\mu_\Im\not=0$ and $\gamma(1+\gamma)>0$, the above formulae
describe a two-dimensional bell-shaped envelope of the plane wave
$\exp(\pm i\varphi)$, exponentially decreasing in all directions in the
$(u,v)$-plane and moving without deformation at the constant velocity
\begin{equation}
\vec V=\left(\matrix{2\mu_\Re\cr 2\lambda_\Re\cr}\right).
\label{3.39}
\end{equation}

The initial position of the soliton is arbitrary, in other words we may
translate the space variables according to the transformation
\begin{equation}
u\rightarrow u-u_0,\qquad v\rightarrow v-v_0,\qquad
u_0,v_0\in\bbbr,                                        \label{3.40}
\end{equation}
and, consequently, the soliton is defined by means of eight real
parameters.

By inserting the value of ${\cal I}(Q^2)$ obtained in (\ref{IQ2}) into
(\ref{timecomponentBT}), rewritten for the gauge
$B(Q,O;\lambda,\mu)$ derived in (\ref{BQOlambdamu}), we get the auxiliary
field $A$
\begin{equation}
A={1\over2}\left(\partial_x-\sigma_3\partial_y\right)\partial_y\log Q^2.
\end{equation}
Finally, from (\ref{gaugeB}) by using the gauge operator
$B(Q,O;\lambda,\mu)$  we obtain the following
eigenfunction relative to the one soliton solution
\beqs
\psi(x,y,k)e^{-ik(\sigma_3x-y)}=\bbbone-{i\over4}\left(
\bmatrix{cc}
\displaystyle\frac{r_Iq}{k-\lambda}&\displaystyle\frac{2q}{k-\mu}\\
\displaystyle-\frac{2r}{k-\lambda}&\displaystyle\frac{q_{II}r}{k-\mu}
\ematrix\right).
\eeqs
The analytic properties of this $\psi$ suggested the special choice for the
Spectral Transform introduced in section 3.

\subsection{Wave soliton solutions of the Hamiltonian DSI}

To obtain soliton solutions of the Hamiltonian DSI equation
(\ref{HamiltonDSI}) we use the elementary B\"acklund gauge in which we keep
the
general space-dependence of the parameters as indicated in
(\ref{alphabetaI}) and (\ref{alphabetaII}).  We start from the
solution $Q=0$ as before but, now, we choose an auxiliary field $A$
of the form $\diag(A_{00}^{(1)}(v,t), A_{00}^{(2)}(u,t))$ where
$A_{00}^{(1)}$ and $A_{00}^{(2)}$ are two arbitrary functions.

The construction of the soliton solution proceeds in the same way as
previously, i.e. by imposing the commutation relation
(\ref{3.19}).  By means of the elementary B\"acklund gauge of the first kind
$B_I(Q_I,0;\lambda)$ defined in (\ref{alphabetaI}) we
obtain the following solution of (\ref{QDSI}) (which does not satisfy
the reduction)
\begin{equation}
Q_I=\left(\matrix{0&0\cr r_I&0\cr}\right),\qquad
A_I=\left(\matrix{A_I^{(1)}(v,t)&0\cr0&A_I^{(2)}(u,t)\cr}\right)
\label{3.41}
\end{equation}
where
\beqs &&r_I(u,v,t)=P(u,t)/D(v,t),
\label{3.42}\\
&&A_I^{(1)}(v,t)=A_{00}^{(1)}(v,t)+2\partial_v^2\log D(v,t),
\label{3.43}\\
&&A_I^{(2)}(u,t)=A_{00}^{(2)}(u,t).
\label{3.44}
\eeqs
The functions $D(v,t)$ and $P(u,t)$ are solutions of the
time dependent Schr\"odinger equations
\beqs
&&iD_t+D_{vv}+A_{00}^{(1)}D=0, \label{3.46}\\
&&iP_t-P_{uu}+A_{00}^{(2)}P=0  \label{3.47}
\eeqs
and
\begin{equation}
\lambda(v,t)=\partial_v\log D(v,t).  \label{3.45}
\end{equation}
By applying now to
the above solution the elementary B\"acklund gauge of the second kind
$B_{II}(Q,Q_I;\mu)$ we obtain the solution
\begin{equation}
Q=\left(\matrix{0&q\cr r&0\cr}\right),\qquad
A=\left(\matrix{A^{(1)}&0\cr0&A^{(2)}\cr}\right)       \label{3.48}
\end{equation}
where
\beqs
&&q=\frac{HD_v-H_vD}{ED+PH/4},\qquad
r=-\frac{PE_u-P_uE}{ED+PH/4},                           \label{3.49}\\
&&A^{(1)}=A_{00}^{(1)}+2\partial_v^2\log(ED+PH/4),
\label{3.50}\\
&&A^{(2)}=A_{00}^{(2)}-2\partial_u^2\log(ED+PH/4).
\label{3.51}
\eeqs
The functions $H(v,t)$ and $E(u,t)$ satisfy the time
dependent Schr\"odinger equations
\beqs
&&iH_t+H_{vv}+A_{00}^{(1)}H=0,
\label{3.52}\\
&&iE_t-E_{uu}+A_{00}^{(2)}E=0.
\label{3.53}
\eeqs
Note that the function $\mu$ and the operator
${\cal I}=\hbox{$1\over2$}\diag(\partial_u^{-1},\partial_v^{-1})$ are not
uniquely determined but are related by the equation
\begin{equation}
\mu(u,t)=-\hbox{$1\over4$}\partial_v^{-1}(rq)-\partial_u\log(ED+PH/4).
\label{3.54}
\end{equation}
The boundary values $A_0^{(1)}$ and
$A_0^{(2)}$ defined in (\ref{HamiltonDSI}) are computed explicitly
from the solutions of the four Schr\"odinger equations and from their
potentials (the inputs $A_{00}^{(1)}$ and $A_{00}^{(2)}$)
\beqs
&&A_0^{(1)} = A_{00}^{(1)}+\partial_v^2\log(ED+PH/4)\vert_{u=-\infty}+
\partial_v^2\log(ED+PH/4)\vert_{u=+\infty},\label{3.55}\\
&&A_0^{(2)} =
A_{00}^{(2)}-\partial_u^2\log(ED+PH/4)\vert_{v=-\infty}-
\partial_u^2\log(ED+PH/4)\vert_{v=+\infty}.
\label{3.55'}
\eeqs

One could get  the same  solution $Q$, $A$ by  applying the
B\"acklund
gauge of first and second kind in the reversed order
$B_I (Q,Q_{II};\lambda )\*B_{II}(Q_{II},0;\mu)$. One gets
\begin{equation}
Q_{II}=\left(\matrix{0&q_{II}\cr0&0\cr}\right),\qquad
A_{II}=\left(\matrix{A_{II}^{(1)}&0\cr0&A_{II}^{(2)}\cr}
\right)         \label{3.56}
\end{equation}
where
\beqs
&&q_{II}(u,v,t)={H(v,t)}/{E(u,t)},
\label{3.57}\\
&&A_{II}^{(1)}(v,t)=A_{00}^{(1)}(v,t),
\label{3.58}\\
&&A_{II}^{(2)}(u,t)=A_{00}^{(u)}(u,t)-2\partial_u^2\log
E(u,t)\label{3.59}
\eeqs
and
\begin{equation}
\mu(u,t)=-\partial_u\log E,
\label{3.60}
\end{equation}
while $\lambda$ is not uniquely determined.  If one
requires that the commutation condition (\ref{3.19}) is satisfied one
gets the  same solution $Q,A$, both $\lambda$ and $\mu$ are determined
according to  the  equations (\ref{3.60}),  (\ref{3.45}) and  the
operator
${\cal I}$ satisfies the equations
\begin{equation}
\partial_u^{-1}qr=qr_I,\qquad \partial_v^{-1}qr=-q_{II}r.
\label{3.61}
\end{equation}
Let  us  consider   the  special   case  of   arbitrary  boundary
values
$A_{00}^{(i)}$ real and moving with constant speed
\beqs
&&A_{00}^{(1)}(v,t)=A_{00}^{(1)}(v+2\phi t) \nonumber\\
&&A_{00}^{(2)}(u,t)=A_{00}^{(2)}(u+2\theta t).
\label{3.62}
\eeqs
If the phases of  the complex functions $D$, $H$ and $E$, $P$
are
chosen to be linear functions it results that
\beqs \bmatrix{ll}
D={\cal D}\exp[-i\delta],\qquad& H=\eta\exp[-i\delta],\\
E={\cal E}\exp[i\epsilon],\qquad& P=\rho\exp[i\epsilon]
\ematrix                        \label{3.63}\eeqs
where
\beqs \bmatrix{ll}
{\cal D}={\cal D}(v+2\phi t),\qquad& \eta=\eta(v+2\phi t),\\
{\cal E}={\cal E}(u+2\theta t),\qquad& \rho=\rho(u+2\theta t)
\ematrix                                \label{3.64}\eeqs
and
\beqs
&&\delta=\phi v+(\phi^2-\phi_0)t+\delta_0,\nonumber\\
&&\epsilon=\theta u+(\theta^2-\theta_0)t+\epsilon_0
    \label{3.65}\eeqs
with $\phi_0$, $\theta_0$, $\delta_0$, $\epsilon_0$ real constants.
The  real   functions ${\cal D}, \eta$ and ${\cal E}$, $\rho$
satisfy the Schr\"odinger equations
\begin{equation}
\bmatrix{c}
{\cal D}_{vv}+(A_{00}^{(1)}-\phi_0){\cal D}=0,\\
\eta_{vv}+(A_{00}^{(1)}-\phi_0)\eta=0
\ematrix                                             \label{07}
\end{equation}
and
\begin{equation}
\bmatrix{c}
{\cal E}_{uu}-(A_{00}^{(2)}+\theta_0){\cal E}=0,\\
\rho_{uu}-(A_{00}^{(2)}+\theta_0)\rho=0.
\ematrix                                             \label{08}
\end{equation}
The solution $Q$ is given by the following formulae
\begin{equation}
q=\frac{W(\eta,{\cal D})}{{\cal
ED}+\rho\eta/4}\exp[-i(\epsilon+\delta)],\qquad
r=-\frac{W(\rho,{\cal E})}{{\cal
ED}+\rho\eta/4}\exp[i(\epsilon+\delta)]                \label{3.68}
\end{equation}
where $W$ is  the  wronskian  operator. The  field $A$ is given by
the
formulae
\begin{equation}
\bmatrix{c}
A^{(1)}=A_{00}^{(1)}+
2\partial_v^2\log({\cal ED}+\rho\eta/4),\\
A^{(2)}=A_{00}^{(2)}-
2\partial_u^2\log({\cal ED}+\rho\eta/4)
\ematrix                                             \label{3.69}
\end{equation}
and its boundary values by
\begin{equation}
\bmatrix{c}
A_0^{(1)}=A_{00}^{(1)}+
\partial_v^2\log({\cal ED}+\rho\eta/4)\vert_{u=-\infty}+
\partial_v^2\log({\cal ED}+\rho\eta/4)\vert_{u=+\infty},\\
A_0^{(2)}=A_{00}^{(2)}-
\partial_u^2\log({\cal ED}+\rho\eta/4)\vert_{v=-\infty}-
\partial_u^2\log({\cal ED}+\rho\eta/4)\vert_{v=+\infty}.
\ematrix                                             \label{3.70}
\end{equation}
The reduced case $r=\sigma_0\bar q$ is
simply obtained
by
requiring that, for $a\in\bbbr$,
\beqs
W(\eta,{\cal D})&=&2a \label{09}\\
W(\rho,{\cal E})&=&-2\sigma_0a.\label{010}
\eeqs
     In conclusion  we obtain,  in the  reduced  case, a  solution of  the
DSI
equation depending on two arbitrary real functions $A^{(1)}_{00}$
 and $A^{(2)}_{00}$.

     The one soliton solution exponentially  decaying in the plane
can be  obtained from this general  solution
by choosing
\begin{equation} \bmatrix{ll}
A_{00}^{(1)}\equiv A_{00}^{(2)}\equiv0,& \lambda_0^2=\phi_0,\qquad
\mu_0^2=\theta_0,
\\ {\cal D}=\exp[\lambda_0(v+2\phi t)],\qquad &
\eta=h_0\cosh[\lambda_0(v+2\phi t)],\\
{\cal E}=\exp[-\mu_0(u+2\theta t)],\qquad &
\rho=p_0\cosh[\mu_0(u+2\theta t)]
\ematrix
\label{3.72}
\end{equation}
where $\lambda_0$, $\mu_0$, $h_0$, $p_0$ are real
constants satisfying the constraints $\sigma_0\mu_0p_0=\lambda_0h_0$
and $h_0p_0>0$.  However, we are here interested in getting solutions
with identically zero boundary values
\begin{equation}
A_0^{(1)}\equiv
A_0^{(2)}\equiv 0
\label{3.73}
\end{equation}
in the
reduced case $r= \sigma_0\overline{q}$.  Because this case has been
     shown to be integrable in the Hamiltonian sense we call it, for
brevity, the Hamiltonian case.

     It is convenient to introduce two functions
$\eta_0(v+2\phi t)$
and $\rho_0(u+2\theta t)$
 defined as follows
\begin{equation}
A_{00}^{(1)}=2\partial_v^2\log\eta_0,\qquad
A_{00}^{(2)}=-2\partial_u^2\log\rho_0.
\label{04}
\end{equation}
     Then, to get a solution of the Hamiltonian case we have to
solve
the   complicated    non   linear    system    of   coupled    equations
for ${\cal D}$, $\eta$, $\eta_0$ and ${\cal E}$, $\rho$, $\rho_0$
\beqs
&&\left.\partial_v^2\log \left[\eta_0{\cal
D}\left(1+\frac{\rho\eta}{4{\cal
ED}}\right)\right]\right|_{u=-\infty}+ \left.\partial_v^2\log
\left[\eta_0{\cal D}\left(1+\frac{\rho\eta}{4{\cal ED}}\right)\right]
\right|_{u=+\infty}=0, \label{05}\\
&&\left.\partial_u^2\log\left[\rho_0{\cal
E}\left(1+\frac{\rho\eta}{4{\cal ED}}\right)\right]
\right|_{v=-\infty}+ \left.\partial_u^2\log\left[\rho_0{\cal E}\left(1+
\frac{\rho\eta}{4{\cal ED}}\right)\right] \right|_{v=+\infty}=0.
\label{06}
\eeqs
Therefore we have to solve the non linear system of
coupled equations (\ref{07}), (\ref{08}), (\ref{05}) and (\ref{06})
with the constraints (\ref{09}) and (\ref{010}).  We add to these
equations the additional constraints
\beqs
&&\lim_{u\to\pm\infty}\frac{\rho}{\cal E}=
-2(\rho_1\pm\rho_2),\label{011}\\
&&\lim_{v\to\pm\infty}\frac{\eta}{\cal D}=
-2(\eta_1\pm\eta_2)
\label{012}
\eeqs
where $\rho_i$ and $\eta_i$ are real constants to
be determined. These requirements allow us to decouple equations
(\ref{05}), (\ref{07}) and (\ref{09}) from equations (\ref{06}),
(\ref{08}) and (\ref{010}). Once having found a special solution $\cal
D$, $\eta$, $\eta_0$ of the first group of equations and a solution
$\cal E$, $\rho$, $\rho_0$ of the second group one has to verify that
they satisfy, respectively, the requirements (\ref{011}) and
(\ref{012}).

Let us first consider the equations (\ref{05}), (\ref{07}), (\ref{09})
and (\ref{011}). From (\ref{05}) and (\ref{011}) we get, by using the
indeterminacy in the definition of $\eta_0$ in (\ref{04}),
\begin{equation}
\eta_0^2 \left[\left({\cal
D}-{1\over2}\rho_1\eta\right)^2-
\left({1\over2}\rho_2\eta\right)^2\right]=1.  \label{013}
\end{equation}
If we
express $\cal D$ in terms of a new function $\alpha(v+\phi t)$ as
follows
\begin{equation}
{\cal D}={1\over2}\rho_1\eta+
\sigma'{1\over2}\rho_2\eta\coth\alpha, \qquad
{\sigma'}^{2}=1,
\label{014}
\end{equation}
we obtain
\beqs
&&\eta=\frac{2\sinh\alpha}{\rho_2\eta_0},\label{015}\\
&&{\cal D}= \frac{\rho_1\sinh\alpha+\sigma'\rho_2\cosh\alpha}{\rho_2\eta_0}
\label{016}
\eeqs
where $\alpha$ and $\eta_0$ are to be determined by requiring that
(\ref{07}) and (\ref{09}) are satisfied. It results that $\eta_0$
decouples from $\alpha$
\begin{equation}
\partial_v^2\eta_0+a^2\rho_2^2\eta_0^5-\phi_0\eta_0=0
\label{017}
\end{equation}
and $\alpha$ can be determined in terms of $\eta_0$
\begin{equation}
 \partial_v\alpha=
\sigma'a\rho_2\eta_0^2.
\label{018}
\end{equation}
For solving the equations (\ref{06}), (\ref{08}) and (\ref{010}) we
introduce, in an analogous way, a new function $\beta(u+2\theta t)$ as
follows
\begin{equation}
{\cal E}={1\over2}\eta_1\rho+\sigma''{1\over2}\eta_2\rho\coth\beta
,\qquad {\sigma''}^{2}=1 .
\label{3.78}
\end{equation}
We get
\beqs
&&\rho=\frac{2\sinh\beta}{\eta_2\rho_0},\label{020}\\
&&{\cal E}=
\frac{\eta_1\sinh\beta+\sigma''\eta_2\cosh\beta}{\eta_2\rho_0}
\label{021}
\eeqs
where $\rho_0$ and $\beta$ are determined by the equations
\beqs
&&\partial_u^2\rho_0+a^2\eta_2^2\rho_0^5-\theta_0\rho_0=0,
\label{022}\\
&&\partial_u\beta=
-\sigma_0\sigma''a\eta_2\rho_0^2.
\label{023}
\eeqs
Finally one has to verify that the following consistency conditions are
satisfied
\beqs
&&\lim_{v\to\pm\infty}(\rho_1+\sigma'\rho_2
\coth\alpha)=-(\eta_1\pm\eta_2)^{-1},\label{024}\\
&&\lim_{u\to\pm\infty}(\eta_1+\sigma''\eta_2
\coth\beta)=-(\rho_1\pm\rho_2)^{-1}.
\label{025}
\eeqs
The solution $q$ can be written as
\begin{equation}
q=\frac {2a\rho_2\eta_2\rho_0\eta_0\exp[-i(\epsilon+\delta)]}
{\sinh\alpha\sinh\beta+(\rho_1\sinh\alpha+\sigma'\rho_2\cosh\alpha)
(\eta_1\sinh\beta+\sigma''\eta_2\cosh\beta)}.
\label{026}
\end{equation}
The ordinary differential equations (\ref{017}) and (\ref{022}) for
$\eta_0$ and $\rho_0$ can be explicitly integrated in terms of
elementary or classical transcendental functions and, consequently, it
is easy to verify the consistency conditions (\ref{024}) and
(\ref{025}).

For the sake of definiteness we consider two cases
\beqs
\bmatrix{lllll}
\mbox{(i)}&\partial_v\eta_0\equiv0,\qquad
&\partial_u\rho_0\equiv0,\qquad&
\phi_0=\lambda^2_0>0,\qquad&\theta_0=\mu_0^2>0;\\
\mbox{(ii)}&\partial_v\eta_0\not\equiv0,\qquad
&\partial_u\rho_0\not\equiv0,\qquad&
\phi_0<0,\qquad&\theta_0<0.
\ematrix
\eeqs
In the case (i)
\begin{equation}
\eta_0^2=\frac{\sigma'\lambda_0}{a\rho_2},\qquad
\rho^2_0=-\frac{\sigma_0\sigma''\mu_0}{a\eta_2}
\label{027}
\end{equation}
and the consistency conditions are satisfied for
\begin{equation}
\sigma'\sigma''\lambda_0\mu_0>0,
\qquad\rho_1\pm\sigma'\sgn(\lambda_0)\rho_2= -{1\over \eta_1\pm\eta_2}.
\label{028}
\end{equation}
If we choose for instance $\sigma''\lambda_0>0$ we get the infinite wave
\begin{equation}
q(u,v,t)=2|\lambda_0\mu_0|^{1/2}\frac
{\exp[-i(\epsilon+\delta)]}{\cosh\xi}
\label{029}
\end{equation}
where
\beqs
&&\xi=\mu_0(u+2\theta t-u_0)-\lambda_0(v+2\phi t-v_0),\nonumber\\
&&\epsilon=\theta u+(\theta^2-\mu_0^2)t+\epsilon_0,\nonumber\\
&&\delta=\phi v+(\phi^2-\lambda_0^2)t+\delta_0.
\label{030}
\eeqs
In case (ii) equation (\ref{017}) can be integrated once to the
equation
\begin{equation}
(\partial_v \eta_0)^2+{1\over3}a^2\rho^2_2\eta_0^6-\phi_0\eta_0^2-
\phi_{00}=0
\label{031}
\end{equation}
with $\phi_{00}$ an arbitrary constant that we choose less than zero.
Its general solution $\eta_0(v+2\phi t)$ can be expressed in terms of
the Weierstrass elliptic function $\wp(v+2\phi t;g_2,g_3)$ with
invariants
\begin{equation}
g_2={4\over3}\phi_0^2,\qquad
g_3={4\over3}a^2\rho^2_2\phi^2_{00}-{8\over27}\phi_0^3
\label{032}
\end{equation}
and negative discriminant
\begin{equation}
\Delta(g_2,g_3)=g_2^3-27g_3^2=-48a^2\rho_2^2\phi_{00}
\left(a^2\rho_2^2\phi_{00}^2-{4\over9}\phi_0^3\right)
\label{033}
\end{equation}
according to the formula
\begin{equation}
\eta_0^2(v+2\phi t)=\frac{\phi_{00}}{{\wp}(v+2\phi
t;g_2,g_3)-\phi_0/3}.
\label{034}
\end{equation}
Note that $\eta_0^2(v)$ for real $v$ is always regular and that
there exists a pure imaginary $v_0$ such that
\begin{equation}
\wp(v_0;g_2,g_3)={\phi_0\over3}, \qquad \partial_v\wp(v_0;g_2,g_3)=
i{2\over\sqrt{3}}a\rho_2\phi_{00}.
\label{035}
\end{equation}
The function $\alpha(v+2\phi t)$ obtained by integrating (\ref{018})
(the invariants $g_2$ and $g_3$ are omitted and $\alpha_0$ is a
constant; $\zeta$ and $\sigma$ are the $\zeta$-- and
$\sigma$--Weierstrass functions)
\begin{equation}
\alpha(v+2\phi t)=\frac{\sigma'a\rho_2\phi_{00}}
{\partial_v{\wp}(v_0)}\left[2(v+\phi t){\zeta}(v_0)+
\log\frac{\sigma(v+2\phi t-v_0)}{\sigma(v+2\phi t+v_0)}\right]
+\alpha_0
\label{036}
\end{equation}
results to be real and to behave for large $u$ as follows
\begin{equation}
\alpha(v+2\phi t)\longrightarrow
2\sigma'a\rho_2\phi_{00}\frac{\zeta(v_0)}{\partial_v\wp(v_0)}v.
\label{037}
\end{equation}
Analogously we get for $\rho_0^2(u+2\theta t)$ ($\theta_0<0$)
\begin{equation}
\rho_0^2(u+2\theta t)=\frac{\theta_{00}}{{\wp}(u+2\theta
t;h_2,h_3)-\theta_0/3}
\label{038}
\end{equation}
where the Weierstrass function $\wp(u;h_2,h_3)$ has invariants
\begin{equation}
h_2={4\over3}\theta_0^2,\qquad
h_3={4\over3}a^2\eta^2_2\theta^2_{00}-{8\over27}\theta_0^3,
\label{039}
\end{equation}
negative discriminant
\begin{equation}
\Delta(h_2,h_3)=h_2^3-27h_3^2=-48a^2\eta_2^2\theta_{00}
\left(a^2\eta_2^2\theta_{00}^2-{4\over9}\theta_0^3\right)
\label{040}
\end{equation}
and
\begin{equation}
\wp(u_0;h_2,h_3)={\theta_0\over3}, \qquad
\partial_u\wp(u_0;h_2,h_3)= i{2\over\sqrt{3}}a\eta_2\theta_{00}
\label{041}
\end{equation}
with $u_0$ pure imaginary.

The function $\beta(u+2\theta t)$ is given by
\begin{equation}
\beta(u+2\theta t)=\frac{\sigma''\sigma_0a\eta_2\theta_{00}}
{\partial_u{\wp}(u_0)}\left[2(u+\theta t){\zeta}(u_0)+
\log\frac{\sigma(u+2\theta t-u_0)}{\sigma(u+2\theta t+u_0)}\right]
+\beta_0
\label{042}
\end{equation}
(invariants $h_2$ and $h_3$ are omitted), which is real and has the
following behaviour at large $u$
\begin{equation}
\beta(u+2\theta t)\longrightarrow -2\sigma_0\sigma''a\eta_2\theta_{00}
\frac{\zeta(u_0)}{\partial_u\wp(u_0)}u.
\label{043}
\end{equation}
The consistency conditions (\ref{024}) and (\ref{025}) are satisfied
for
\begin{equation}
\sigma'\sigma''\rho_2\theta_{00}\phi_{00}
\frac{\zeta(v_0)\zeta(u_0)}{\partial_v\wp(v_0) \partial_u\wp(u_0)}>0
\label{044}
\end{equation}
and
\begin{equation}
\rho_1\pm\sgn\left(a\rho_2\phi_{00}
\frac{\zeta(v_0)}{\partial_v\wp(v_0)}\right)\rho_2=
{1\over\eta_1\pm\eta_2}.
\label{045}
\end{equation}
By inserting these values and functions in the equation (\ref{026}) for
$q$ we get an infinite wave with a periodically modulated amplitude.

\section{Solitons via the Spectral Transform}
\setcounter{equation}{0}

\subsection{The Spectral
Transform of the Kadomtsev--Petviashvili~I equation}

It turns out that the Spectral Transform for the Kadomtsev--Petviashvili~I
equation plays a relevant role in the study of the DSI equation.
Therefore, this section is dedicated to the main properties
of this Spectral Transform that are of interest in this respect. Specifically,
we will derive the multi--wave-soliton solution of the
Kadomtsev--Petviashvili~I equation and the
orthogonality relations for the eigenfunctions of the associated
spectral problem.

We consider the Kadomtsev--Petviashvili equation in its variant
(called KPI)
\begin{equation}
(u_t-6uu_x+u_{xxx})_x=3u_{yy}
\end{equation}
with $u=u(x,y,t)$ real. Its Spectral Transform
is defined via the associated ``time''
(the space variable $y$ plays here the role of time) dependent
Schr\"odinger equation
\begin{equation}
-i\Phi_y+\Phi_{xx}-u\Phi=0.
\label{spectraleqKP}
\end{equation}

The spectral parameter $k$ is introduced by requiring that
\begin{equation}
\Phi(x,y,k)e^{ikx-ik^2y}=1+O\left({1\over k}\right),\qquad k\rightarrow\infty.
\label{asympKP}
\end{equation}

We consider, first, the case in which $u$ is going sufficiently fast to
zero at large distances in the $(x,y)$ plane. Then, the eigenfunction $\Phi$
can be chosen to be bounded in the $(x,y)$ plane and sectionally holomorphic
in the complex $k$--plane. More precisely, $\Phi$ that is called the
Jost solution is analytic in the upper and in the lower half plane and
its boundary values $\Phi^{\pm}$ on the two sides $\pm\mbox{Im}\, k>0$
of the real $k$--axis are given by the integral equation
\begin{equation}
\Phi^\sigma(x,y,k)=\int\!\! dp\, e^{-ipx+ip^2y}R^\sigma(y,k,p),
\qquad\sigma=\pm,
\label{intKP}
\end{equation}
where
\beqs
& &R^\sigma(y,k,p)=\delta(k-p)-{\sigma\over{2\pi i}}\sgn(p-k)\nonumber\\
& &\times\int\!\! d\eta\,\theta\big(\sigma
(y-\eta)(p-k)\big)\int\!\! d\xi\,e^{ip\xi-ip^2\eta}u(\xi,\eta)\Phi^\sigma
(\xi,\eta,k).\label{Rsigma}
\eeqs
When it is not differently indicated the integration is performed all along
the real axis from $-\infty$ to $+\infty$.

The Spectral Transform $F(k,l)$ of the potential $u(x,y)$ is
defined as the measure of the departure from analyticity of the Jost
solution $\Phi$
\begin{equation}
{{\partial\Phi}\over{\partial\overline{k}}}=\int\!\!\int\!\!dl\wedge
d\overline{l}
\,\Phi(l)F(k,l)
\label{DbarKP}
\end{equation}
where
\begin{equation}
{{\partial}\over{\partial\overline{k}}}\equiv{1\over2}\left(
{\partial\over{\partial k_\Re}}+i{\partial\over{\partial k_\Im}}\right),
\qquad
k_\Re=\mbox{Re}\, k,\,\,k_\Im=\mbox{Im}\, k
\end{equation}
is the so called $\overline{\partial}$-derivative.

The main quantity to study in order to get complete information
on the Spectral Transform of the KPI equation is the scalar product
\begin{equation}
\bigl<\Phi^\sigma(k),\Phi^{\sigma'}(p)\bigr>\equiv{1\over{2\pi}}\int\!\!dx\,
\Phi^\sigma(x,y,k)\overline{\Phi^{\sigma'}(x,y,p)}.\label{scalarpr}
\end{equation}
By inserting here the integral equation for $\Phi^\sigma$ and
$\Phi^{\sigma'}$ we get
\begin{equation}
\bigl<\Phi^\sigma(k),\Phi^{\sigma'}(p)\bigr>=\Bigl({\bf R}^\sigma(y)
{{\bf R}^{\sigma'}}^{\dag}(y)\Bigr)(k,p),
\end{equation}
where $R^\sigma(y,k,p)$ is considered as the kernel of an integral operator
${\bf R}^\sigma (y)$, ${R^\sigma}^{\dag}(y,k,p)=\overline{R^\sigma(y,p,k)}$
is the kernel of the adjoint operator ${{\bf R}^\sigma}^{\dag}(y)$ and
\begin{equation}
\Bigl({\bf R}^\sigma(y){{\bf R}^{\sigma'}}^{\dag}(y)\Bigr)(k,p)=
\int\!\!dl\,R^\sigma(y,k,l)\overline{R^{\sigma'}(y,p,l)}
\end{equation}
is the kernel of the product ${\bf R}^\sigma(y){{\bf R}^{\sigma'}}^{\dag}(y)$.
By differentiating  the scalar product (\ref{scalarpr})
with respect to $y$ and by using
 (\ref{spectraleqKP})
for $\Phi^\sigma$ and $\Phi^{\sigma'}$ one proves that it is
$y$ independent. For exploiting this information it is convenient
to introduce
\begin{equation}
R_\pm^\sigma(k,p)=\lim_{y\rightarrow\pm\infty}R^\sigma(y,k,p).
\label{Rpm}
\end{equation}
They are kernels of the triangular integral operators ${\bf R}_{\pm}^\sigma$
whose explicit expressions
\beqs
R_\pm^\sigma(k,p)&=&\delta(k-p)\mp\vartheta\bigl(\mp\sigma(k-p)\bigr)r^\sigma
(k,p),\label{Rfordecaying}\\
r^\sigma(k,p)&=&{1\over{2\pi i}}\int\!\! d\eta\!\!
\int\!\! d\xi\,e^{ip\xi-ip^2\eta}u(\xi,\eta)\Phi^\sigma
(\xi,\eta,k),\quad\sigma=\pm,\label{rPhi}
\eeqs
are obtained by inserting into (\ref{Rsigma}) the identity
\begin{equation}
\sigma\sgn(p-k)\vartheta\bigl(\sigma(y-\eta)(p-k)\bigr)=
\mp\vartheta\bigl({\mp}
(y-\eta)\bigr)\pm\vartheta\bigl({\mp}\sigma(k-p)\bigr).
\label{thetaidentity}
\end{equation}
Then, the $y$ independence of the scalar product implies that
\begin{equation}
\bigl<\Phi^\sigma(k),\Phi^{\sigma'}(p)\bigr>=\Bigl({\bf R}^\sigma_+
{{\bf R}^{\sigma'}_+}^{\dag}\Bigr)(k,p)=\Bigl({\bf R}^\sigma_{-}
{{\bf R}^{\sigma'}_{-}}^{\dag}\Bigr)(k,p).
\label{characterizationI}
\end{equation}
In the case $\sigma'=-\sigma$ the two operators
${\bf R}^\sigma_+{{\bf R}^{-\sigma}_+}^{\dag}$ and
${\bf R}^\sigma_{-}{{\bf R}^{-\sigma}_{-}}^{\dag}$
are, respectively, lower and upper triangular or upper and lower triangular
according to the sign, positive or negative, of $\sigma$. Consequently,
the equality in (\ref{characterizationI}) implies that
\begin{equation}
\label{characterizationII}
{\bf R}^\sigma_ +{\bf R^{-\sigma}_+}^{\dag}={\bf R}^\sigma_{-}{{\bf
R}^{-\sigma}_{-}}^{\dag}={\bf I},
\end{equation}
where the kernel of the unity operator is $I(k,p)=\delta(k-p)$, so that
$\Phi^\sigma$ and $\Phi^{-\sigma}$ are orthogonal
\begin{equation}
\label{scalarpr'}
\bigl<\Phi^\sigma(k),\Phi^{-\sigma}(p)\bigr>=\delta(k-p).
\end{equation}

In view of the relevant role played by ${\bf R}^\sigma_{\pm}$ it is
convenient,
by using again the identity (\ref{thetaidentity}) and the definition
(\ref{Rpm}), to recast the integral equation (\ref{intKP}) in the form
\beqs
& &\Phi^\sigma(x,y,k)=\int\!\! dp\,e^{-ipx+ip^2y}R^\sigma_\pm(k,p)\nonumber\\
\label{integralKP}
& &-{1\over{2\pi i}}\int\!\! dp\!\!\int_{\pm\infty}^y\!\! d\eta\!\!
\int\!\! d\xi\,e^{ip(\xi-x)-ip^2(\eta-y)}u(\xi,\eta)\Phi^\sigma(\xi,\eta,k).
\eeqs

Let us now turn to the eigenfunctions $\Psi_\pm$ solution of
the integral equation
\begin{equation}
\Psi_\pm(x,y,k)=e^{-ikx+ik^2y}
-{1\over{2\pi i}}\int\!\! dp\!\!\int_{\pm\infty}^y\!\! d\eta\!\!
\int\!\! d\xi\,e^{ip(\xi-x)-ip^2(\eta-y)}u(\xi,\eta)\Psi_\pm
(\xi,\eta,k).\label{Psi}
\end{equation}
{}From (\ref{integralKP}) we see that
\begin{equation}
\Phi^\sigma={\bf R}^\sigma_\pm\Psi_\pm.
\label{PhiPsi}
\end{equation}
As we know by (\ref{characterizationII}) the operators
${\bf R^{-\sigma}_\pm}^{\dag}$ are right inverse of the ${\bf R}^\sigma_\pm$.
Let the potential $u$ be such that these operators are both sides mutually
inverse,
i.e. let be besides (\ref{characterizationII})
\begin{equation}
\label{characterizationIII}
{{\bf R}_+^{-\sigma}}^{\dag}{\bf R}^\sigma_+={{\bf R}_-^\sigma}^{\dag}{\bf
R}^{-\sigma}_-={\bf I}.
\end{equation}
Then  the relation (\ref{PhiPsi}) can be rewritten as
\begin{equation}
\label{PsiRPhi}
\Psi_\pm={{\bf R}^{-\sigma}_\pm}^{\dag}\Phi^\sigma.
\end{equation}
Due to the defining integral equation (\ref{Psi}) the
solutions $\Psi_\pm$ are $\sigma$ independent. Consequently  equation
(\ref{PsiRPhi}) yields
\begin{equation}
\label{sigmaindependence}
{{\bf R}_\pm^{-\sigma}}^{\dag}\Phi^\sigma={{\bf
R}_\pm^\sigma}^{\dag}\Phi^{-\sigma}
\end{equation}
and, by using again (\ref{characterizationII}),
\begin{equation}
\label{jumpF}
\Phi^\sigma=\hbox{\bfcal F}^{\,-\sigma}\Phi^{-\sigma},
\end{equation}
where
\begin{equation}
\label{definitionF}
\hbox{\bfcal F}^{\,-\sigma}={\bf R}_\pm^\sigma{{\bf R}_\pm^\sigma}^{\dag}.
\end{equation}
{}From (\ref{characterizationI}) and (\ref{definitionF}) we have directly that
\begin{equation}
\bigl<\Phi^\sigma(k),\Phi^\sigma(p)\bigr>={\cal F}^{\,-\sigma}(k,p).
\end{equation}
It is convenient to separate the $\delta$ distribution contained in
the kernel of $\hbox{\bfcal F}^{\,\sigma}$ by writing
\begin{equation}
\label{definitionf}
{\cal F}^\sigma(k,p)=\delta(k-p)-\sigma f^{\sigma}(k,p).
\end{equation}
Then, equations (\ref{jumpF}) and (\ref{definitionF}) read
\begin{equation}
\label{jumpf}
\Phi^+(x,y,k)-\Phi^{-}(x,y,k)=\int\!\! dl\,f^\sigma(k,l)\Phi^\sigma(x,y,l),
\end{equation}
\begin{equation}
\label{relationfr}
f^{-\sigma}(k,p)=r^\sigma(k,p)\vartheta(k-p)+\overline{r^\sigma(p,k)}
\vartheta(p-k)
+\sigma\!\!\int_{-\infty}^k\!\! dl\,\vartheta(p-l)r^\sigma(k,l)
\overline{r^\sigma(p,l)}.
\end{equation}

By recalling that on the real $k$-axis
\begin{equation}
{{\partial\Phi}\over{\partial\overline{k}}}(k)={i\over2}\bigl(\Phi^+(k)
-\Phi^-(k)\bigr)\delta(k_\Im)
\end{equation}
we get for the Spectral Transform of $u$ defined in
(\ref{DbarKP})
\begin{equation}
F(k,l)={i\over 2}f^\sigma(k,l)\delta(k_\Im).
\end{equation}

Formula (\ref{relationfr}) solves the direct spectral problem furnishing an
explicit expression of the Spectral Transform in
terms of $u$ and of the Jost solution $\Phi^\sigma$ via $r^\sigma(k,p)$.

We note that the operators $\hbox{\bfcal F}^{\,\sigma}$($\sigma=\pm$),
which yield the Spectral Transform of $u$, are selfadjoint
\begin{equation}
{\hbox{\bfcal F}^{\,\sigma}}^{\dag}=\hbox{\bfcal F}^{\,\sigma}
\label{characterizationFI}
\end{equation}
and one is the inverse of the other
\begin{equation}
\hbox{\bfcal F}^{\,\sigma}\hbox{\bfcal F}^{\,-\sigma}={\bf I}.
\label{characterizationFII}
\end{equation}
These two equations can be considered as the characterization equations for
the
Spectral Transform $\hbox{\bfcal F}^{\,\sigma}$ (in the sense that they
determine the
class of admissible spectral data) if
$\hbox{\bfcal F}^{\,\sigma}$ is the product, as in
(\ref{characterizationII}),
of two
operators ${\bf R}^\sigma_\pm$ which have the triangular form indicated in
(\ref{Rfordecaying}). Because the characterization equations are more simply
expressed in terms of $\hbox{\bfcal F}^{\,\sigma}$ we prefer, in the
following,
to define $\hbox{\bfcal F}^{\,\sigma}$ as the Spectral Transform of $u$.

The reconstruction of the potential $u$ can be
performed starting from the Spectral Data $\hbox{\bfcal F}^{\,\sigma}$.
First,
one
solves the singular Fredholm equation
\begin{equation}
\Phi^\sigma(x,y,k)=e^{-ikx+ik^2y}
-{\sigma'\over 2\pi i}\int\!\!dq\,{e^{-i(k-q)x+i(k^2-q^2)y}
\over q-k-i0\sigma}\!\!\int\!\!dp\,[\hbox{\bfcal F}^{\,\sigma'}-{\bf I}](q,p)
\Phi^{\sigma'}(x,y,p).\quad
\label{CauchyGreen}
\end{equation}
The obtained $\Phi^\sigma$, by the Cauchy-Green theorem, solves the non local
Riemann-Hilbert problem expressed by the $\overline{\partial}$-equation
(\ref{DbarKP}) (or its equivalent form (\ref{jumpF})) and satisfies
the asymptotic requirement (\ref{asympKP}). Secondly, one expands
in powers of $1/k$ the fraction $1/(q-k-i0\sigma)$ in (\ref{CauchyGreen}) and
inserts the obtained asymptotic expansion in (\ref{spectraleqKP}).
We have
\begin{equation}
u(x,y)=-\sigma'{1\over \pi} {\partial \over \partial x}
\int\!\! dq\!\!\int\!\! dp\,[\hbox{\bfcal F}^{\,\sigma'}-{\bf I}](q,p)
\Phi^{\sigma'}(x,y,p)e^{iqx-iq^2y}.
\label{inverse}
\end{equation}

In the following section we are interested also in potentials $u$
describing $N$ interacting wave solitons. Then, $u$ goes to a constant
along $N$ directions in the $(x,y)$ plane and the Green function in the
integral equation (\ref{intKP}) has to be corrected in order to avoid
divergences. The theory of the Spectral Transform does not result to be
substantially
different from the theory in the case of $u$ vanishing at large
distances.  However, there are many subtle technical difficulties to
handle which are out of the scope of the present paper. Therefore,
we restrict the discussion to the main points and we refer for details
to the published papers by two of the authors of this paper (M.~B.
and F.~P.) and by A.~Pogrebkov and M.~Polivanov.

The Jost solution $\nu(x,y,k)=\Phi(x,y,k)\exp[ikx-ik^2y]$ in the complex
$k$--plane
can be defined as the solution of the following integral equation
\beqs
& &\nu(x,y,k)=1+{1\over (2\pi)^2} \int\!\!\int\!\! dpdq\,{1\over
q-(p-i0)(p+2k)}\nonumber\\
& &\times\int\!\!\int\!\! d\eta d\xi\,\nu(\xi,\eta,k) u(\xi,\eta)
e^{ip(\xi-x)-iq(\eta-y)}
\label{Jost}
\eeqs
where the integrations must be done in the order indicated from the right
to the left. For a potential $u$ going sufficiently fast to zero for large
$x^2+y^2$ the order is unessential and one can explicitly perform, first,
the integration over $q$ recovering the integral equation (\ref{intKP}).

The integral equation (\ref{Jost}) contains as a special case the
unidimensional case. In fact if $u(x,y)=\widetilde u(x)$ by introducing
\begin{equation}
\widetilde\Phi(x,k)=e^{-ik^2y} \Phi(x,y,k)
\end{equation}
one recovers the stationary Schr\"odinger equation
\begin{equation}
\widetilde\Phi_{xx}(x,k) +\big(k^2-\widetilde u(x)\big)\widetilde\Phi(x,k)=0.
\end{equation}
Moreover, in (\ref{Jost}), once integrated over $\eta$, the integration over
$\xi$ and $p,q$ can be interchanged and one obtains by computing the limit
$k_{\Im}\to0$ on the two sides of the real k-axis the familiar integral
equations
\beqs
\widetilde\Phi^+(x,k)&=&e^{-ikx}\nonumber\\
& & { }+\int\!\! d\xi {e^{ik\vert x-\xi\vert}\over 2i(k+i0)}\widetilde
u(\xi)\widetilde\Phi^+(\xi,k)\\
\widetilde\Phi^-(x,k)&=&e^{-ikx}\nonumber\\& &{ }-\int\!\! d\xi\,\theta (\xi-x)
{\sin[k(x-\xi)]\over k-i0} \widetilde u(\xi)\widetilde\Phi^-(\xi,k)
\eeqs
defining the sectionally meromorphic Jost solution $\widetilde\Phi$ of the
stationary
Schr\"odinger equation.

In the general bidimensional case, as in the previous case ($u$ vanishing
at infinity), the main quantities to study are the scalar products of the
Jost solutions. They result to be still $y$ independent and, therefore, formula
(\ref{characterizationI}) remains valid. The most difficult point is the
computation
of the limits $R_\pm^\sigma$ and of the characterization equations. It results
that formula (\ref{Rfordecaying}) must be changed as follows
\begin{equation}
R^\sigma_\pm(k,p)=\delta(k-p)\big(1+Z^\sigma_\pm(k)\big)\mp\theta(\mp
\sigma(k-p)) r^\sigma(k,p)
\end{equation}
by adding to the coefficient of the $\delta$ distribution a function
 $Z^\sigma_\pm(k)$ to be determined.

The Jost solutions $\Phi^\sigma$ and $\Phi^{-\sigma}$ are still orthogonal
and this is the main property we will use in the following section.

Let us now proceed and consider the $N$-soliton solution of the KPI, namely
the potential $u$ characterized by a Spectral Transform containing
exclusively a number of discrete eigenvalues $\mu_n\in\bbbc$
$(n=1,\dots,N).$
For real $u$ this spectrum $f_d(k,l)$ must satisfy the characterization
equation
\begin{equation}
\overline f_d(k,l)=f_d(l,k)
\end{equation}
and therefore its most general form is
\begin{equation}
f_d(k,l)=\sum^N_{n=1} \sum^N_{m=1}
r_{nm}\delta(l-\overline\mu_m)\delta(k-\mu_n)
\end{equation}
with $r_{nm}$ an arbitrary constant hermitian matrix. It is convenient
to parameterize this matrix as follows
\begin{equation}
r_{nm}=2\pi i
\exp[i\overline\mu_m(x_{om}-\overline\mu_my_{om})-i\mu_n(x_{on}-
\mu_n y_{on})]C_{nm}{\mu\llap{$-$}}_{mm}
\end{equation}
where the real
parameters $x_{on}$ and $y_{on}$ fix the initial position of the
$n^{th}$ soliton in the plane,
\begin{equation} {\mu\llap{$-$}}_{nm}\equiv
\overline\mu_n-\mu_m,
\end{equation}
and the complex matrix $C$ satisfies
\begin{equation}
C^2_{nn}=1, \qquad C_{nm}{\mu\llap{$-$}}_{mm}=\overline
C_{mn}{\mu\llap{$-$}}_{nn}.
\end{equation}
For definiteness we choose $C_{nn}=1$ and $\mbox{Im}\,\mu_n>0$.

Because the Jost solution $\Phi(x,y,k)$ in the pure discrete case has only
simple poles at $k=\mu_n$ $(n=1,\dots,N)$ and satisfies the
asymptotic property (\ref{asympKP}) it admits the representation
\begin{equation}
\Phi(x,y,k)=e^{-ikx+ik^2y}\left(1+\sum^N_{n=1} {\varphi_n(x,y)\over k-\mu_n}
e^{\alpha_n}\right)
\end{equation}
where, for  convenience, it has been introduced an exponential factor
$e^{\alpha_n}$
 with
\begin{equation}
\alpha_n=i\mu_n(x-x_{on})-i\mu_n^2(y-y_{on})
\end{equation}
The insertion of this representation into the $\overline\partial-$equation
(\ref{DbarKP}) yields an algebraic equation for the functions $\varphi_n$
\begin{equation}
\label{sumN}
\sum^N_{m=1}{\cal A}_{nm}\varphi_m = -\sum^N_{m=1}
C_{nm}{\mu\llap{$-$}}_{mm}e^{\overline\alpha_m}
\end{equation}
where
\begin{equation}
{\cal A}=1+C\alpha,\qquad\alpha_{nm}={{\mu\llap{$-$}}_{nn} \over
{\mu\llap{$-$}}_{nm}}e^{\overline\alpha_n+\alpha_m},
\end{equation}
while its insertion into (\ref{inverse}) yields the potential $u$
\begin{equation}
u=-2i\partial_x \sum^N_{n=1} \varphi_ne^{\alpha_n}.
\end{equation}

To see that (\ref{sumN}) can be solved for $\varphi_n$ it is sufficient to show
that
${\cal A}$ is positive definite. In the following section in the more
general framework
of the Spectral Transform for the DSI equation we prove that this is true if
the hermitian
matrix $i C_{nm}{\mu\llap{$-$}}_{mm}$ has positive eigenvalues.

By the rule for differentiating a determinant the $N$ soliton solution $u$
can be written in the closed form
\begin{equation}
u=-2\partial^2_x \ln\det{\cal A}
\end{equation}
and the orthogonality relations for the Jost solutions
\begin{equation}
i\!\!\int\!\! dx\,\overline{\Phi^-(x,y,\overline\mu_m)}
\Res\big(\Phi^+(x,y,k),\mu_n\big)=\delta_{mn}
\end{equation}
can be derived.

The potential $u(x,y)$ is going to a constant at large distance along
the directions $x-2\mbox{Re}\,\mu_{n}y=const\,$ and describes $N$
intersecting waves of infinite length.

\subsection{The Spectral
Transform of the Davey-Stewartson I equation}

We consider the DSI equation (\ref{QDSI}) with boundary conditions
defined as in (\ref{ADSI}). The real boundaries $a_0^{(1)}(v,t)$
and $a_0^{(2)}(u,t)$ are assumed to go to zero at large distances in
the $(v,t)$ and $(u,t)$ plane, respectively, with the possible
exception of a finite number of directions along which they are going
to some constants.

According to the usual scheme, in order to linearize the DSI equation,
we have to define the Spectral Transform for the Zakharov--Shabat spectral
problem in the plane
(hyperbolic case)
\begin{equation}
\label{4.1}
T_1\psi\equiv (\partial_x +\sigma_3\partial_y +Q)\psi = 0
\end{equation}
The complex spectral parameter $k$ is introduced by requiring that the
$2\times 2$ matrix Jost solution $\psi$ satisfies the asymptotic property
\begin{equation}
\label{4.2}
\psi (x,y,k)e^{-ik(\sigma_3 x-y)}={\bbbone }+O({1\over k}),
\qquad k\rightarrow \infty
\end{equation}
The Green function of the Zakharov--Shabat
spectral problem can be chosen to be sectionally holomorphic and the
values $\psi^\pm$ of $\psi$ on the two sides $\pm \mbox{Im}\,k>0$ of
the real axis in the $k$-plane are given by the integral equations
\begin{equation}
\label{4.3}
\psi^{\pm}=\psi_0^\pm +G^\pm\psi^\pm
\end{equation}
where
\beqs
\label{4.4}
G^\pm\psi & = & (G_1^\pm\mbox{\boldmath$\psi$}_1,
G^\pm_2\mbox{\boldmath$\psi$}_2)\\
\psi & \equiv & (\mbox{\boldmath$\psi$}_1,\mbox{\boldmath$\psi$}_2)\equiv
\left ( \bmatrix{cc}
\psi_{11}& \psi_{12}\\
\psi_{21}& \psi_{22}
\nonumber
\ematrix\right )
\eeqs
with
\beqs
G_1^\pm&=&{1\over2}\left(
\bmatrix{cc}
0&\int^{-\infty}_u\!du'\,q(u',v)\\
\int^{\pm\infty}_v\!dv'\,r(u,v')&0
\ematrix\right)\nonumber\\
G_2^\pm&=&{1\over2}\left(
\bmatrix{cc}
0&\int^{\mp\infty}_udu'\,q(u',v)\\
\int^{-\infty}_vdv'\,r(u,v')&0
\ematrix\right)
\label{4.5}
\eeqs
and $\psi_0^\pm=\diag(\psi^\pm_{01},\psi^\pm_{02})$ is an arbitrary
solution of the homogeneous part of the Zakharov--Shabat spectral equation
\begin{equation}
\label{4.6}
(\partial_x+\sigma_3\partial_y)\psi^\pm_{0}=0
\end{equation}
It is worth noting for future use that the Green operators $G^\pm$ have
the symmetric property
\begin{equation}
\label{4.7}
G_2^+=G_1^-
\end{equation}
and are $k$ independent.

The $2\times2$ matrix Spectral Transform $R(k,l))$ of $Q(x,y)$ is
defined as the measure of the departure from analyticity of $\psi$
\begin{equation}
\label{4.8}
\frac{\partial}{\partial\overline
k}\psi(x,y,k)=\!\!\int\!\!\!\int_{\bbbc}\!\!dl\wedge
d\overline{l}\,\psi(x,y,l)R(k,l).
\end{equation}

In contrast with the one dimensional case, once chosen the Green
operator, the asymptotic requirement (\ref{4.2}) does not  fix
$\psi_{01}$ and $\psi_{02}$ which are arbitrary functions of $v$ and
$u$, respectively (see (\ref{4.6})). Therefore, for different choices of
$\psi_0$ we get different $\psi$ and consequently via the definition
(\ref{4.8}) different Spectral Transforms $R(k,l)$.

We search for a Spectral Transform that satisfies the following two
fundamental properties:
\begin{itemize}
\item[i)]
its time evolution can be explicitly integrated;
\item[ii)]
the discrete  part of the spectrum corresponds to solitons and the
continuous part to radiation.
\end{itemize}
In order to satisfy the requirement $i)$, in analogy with the
one dimensional case, we fix the time evolution of $\psi$ by requiring
that
\begin{equation}
\label{4.10}
T_2\psi=\psi\Omega(k)
\end{equation}
where $T_2$ is the second Lax operator in (\ref{1.13}) and
$\Omega(k)=-\sigma_3k^2$ is the dispersive function of the DSI
equation. In fact, by applying to both sides of (\ref{4.10}) the
operator $\partial\over\partial\overline k$ and by using the definition
of the Spectral Transform in (\ref{4.8}) we get the linear time evolution
equation for
the Spectral Transform
\begin{equation}
\label{4.11}
iR_t(k,l,t)=R(k,l,t)\Omega(k)-\Omega(l)R(k,l,t)
\end{equation}
which is easily explicitly solved getting
\begin{equation}
\label{4.12}
R(k,l,t)=e^{i\Omega(l)t}R(k,l,0)e^{-i\Omega(k)t}
\end{equation}
By taking the limit ($u\to-\infty$, $v$ fixed) for the $(\;)_{11}$
matrix element of (\ref{4.10}) and the limit ($v\to-\infty$, $u$ fixed)
for the $(\;)_{22}$ element we derive two time dependent Schr\"odinger
equations for $\psi_{01}$ and $\psi_{02}$
\beqs
\nonumber
&&[\partial^2_v+k^2+a_0^{(1)}(v,t)]\psi_{01}(v,t,k)=
-i\partial_t\psi_{01}(v,t,k)\\
&&[\partial^2_u+k^2-a_0^{(2)}(u,t)]\psi_{01}(u,t,k)=
i\partial_t\psi_{02}(u,t,k)
\label{4.13}
\eeqs
These equations together with the asymptotic property
\beqs
\label{4.14}
\psi_0e^{-ik(\sigma_3x-y)}=\bbbone+O({1\over k}),\qquad k\to\infty,
\eeqs
which can be derived from (\ref{4.3}) by using (\ref{4.2}), uniquely
determine $\psi_0$ in terms of the boundary values $a^{(1)}_0$ and
$a^{(2)}_0$.
Precisely, we have
\beqs
\psi_{01}(v,t,k)&=&\Phi^{(1)}(v,t,k)e^{ik^2t}\\
\psi_{02}(u,t,k)&=&\Phi^{(2)}(u,t,k)e^{-ik^2t}
\eeqs
where $\Phi^{(1)}(v,t,k)$ and $\Phi^{(2)}(u,t,k)$ are the Jost
solutions of the time dependent Schr\"odinger equations with space
variables $v$ and $u$ and potentials $a^{(1)}(v,t)$ and
$a^{(2)}(u,t)$, respectively.

The integral equations (\ref{4.3}) for $\psi$ are of Volterra type and
therefore the singularities of $\psi$ in the complex $k$--plane  are
those of $\psi_0$ and of the sectionally holomorphic Green function.
Therefore we need to revisit the Spectral Transform for the time
dependent Schr\"odinger equation. In particular if $a^{(1)}_0$ and
$a^{(2)}_0$ are wave solitons in the plane $(v,t)$ and $(u,t)$ the
eigenfunction $\psi_0$ and consequently $\psi$ have simple poles in
$k$. The existence and the location of poles are uniquely determined by
the boundary values $a^{(1)}_0$ and $a^{(2)}_0$. However, we will see
that the boundaries do not characterize completely the discrete spectrum
and one is left with an unexpected freedom in choosing other
independent parameters.

The continuous part $R_c(k,l)$ of the Spectral Transform $R(k,l)$
measures the discontinuity $\psi_+-\psi_-$ of $\psi$ along the real k
axis and it has the form
\begin{equation}
R_c(k,l)=-{1\over4}\left(
\bmatrix{cc}
\delta(l_\Im+0)&0\\
0&\delta(l_\Im-0)
\ematrix\right)
\left(\bmatrix{cc}
F_1(k,l)&-S_2(k,l)\\
S_1(k,l)&F_2(k,l)
\ematrix\right)\delta(k_\Im)
\label{4.15}
\end{equation}
where we make the choice $\left(\bmatrix{cc}\delta(l_\Im+0)&0\\
0&\delta(l_\Im-0)\ematrix\right)$ instead of the usual one
$\delta(l_\Im+0)\bbbone$ or $\delta(l_\Im-0)\bbbone$ in order to exploit in
the following the
symmetry $G_2^+=G_1^-$ of the Green operator.

Let us introduce the integral operator
\begin{equation}
\label{4.16}
G=(G_1^-,G_2^+).
\end{equation}
{}From the integral equation (\ref{4.3}) we derive directly ($k$ is real,
space time variables are understood)
\begin{equation}
\label{4.17}
(\psi^+-\psi^-)(k)={\cal F}_{c}(k)+G(\psi^+-\psi^-)(k)
\end{equation}
where
\begin{equation}
\label{4.18}
{\cal F}_c(k)=\left(\bmatrix{cc}
(\psi^+_{01}-\psi^-_{01})(k)
&-{1\over2}\!\!\int\!\!du\,q\psi^-_{22}(k)\\[10pt]
{1\over2}\!\!\int\!\!dv\,r\psi^+_{11}(k)&(\psi^+_{02}-\psi^-_{02})(k)
\ematrix\right).
\end{equation}
Here and in the following when it is not differently indicated the
integration is performed all along the real axis from $-\infty$ to
$+\infty$.

By inserting the same integral equation (\ref{4.3}) into the r.h.s. of
(\ref{4.8}), by using the symmetry $G_2^+=G_1^-$ and the $k$
independence of $G^\pm$ and by recalling that on the real k axis
\begin{equation}
\label{4.19}
\frac{\partial\psi}{\partial\overline
k}(k)={i\over2}(\psi^+(k)-\psi^-(k))\delta(k_\Im)
\end{equation}
we get
\begin{equation}
\label{4.20}
(\psi^+-\psi^-)(k)=\widetilde{\cal F}_{c}(k)+G(\psi^+-\psi^-)(k)
\end{equation}
where
\begin{equation}
\label{4.21}
\widetilde{\cal F}_{c}(k)=\!\!\int\!\!dl\left(\bmatrix{cc}
\psi_{01}^-(l)F_1(k,l)&-\psi_{01}^-(l)S_2(k,l)\\[10pt]
\psi^+_{02}(l)S_1(k,l)&\psi^+_{02}(l)F_2(k,l)
\ematrix\right)
\end{equation}
Since the operator $G$ is of Volterra type the related homogeneous
integral equation has only the vanishing solution and, consequently, by
comparing the two integral equations (\ref{4.17}) and (\ref{4.20}) we
get
\begin{equation}
\label{4.22}
{\cal F}_c=\widetilde{\cal F}_{c}.
\end{equation}
{}From this equation we deduce that $F_1$ and $F_2$ are the continuous
component of the Spectral Transform of the boundaries $a^{(1)}_0$ and
$a^{(2)}_0$ and,
by using the orthogonality relations (\ref{scalarpr'})
rewritten for the eigenfunctions $\psi_{01}$ and $\psi_{02}$, we
express explicitly $S_1$ and $S_2$ in terms of $Q$, $\psi$ and $\psi_0$
as follows
\beqs
\nonumber S_1(k,l)&=
&{1\over4\pi}\!\!\int\!\!\!\int\!\!du\,dv\,r(u,v)\psi^+_{11}(u,v,k)
\overline{\psi^-_{02}}(u,l)\\
S_2(k,l)&=
&{1\over4\pi}\!\!\int\!\!\!\int\!\!du\,dv\,q(u,v)\psi^-_{22}(u,v,k)
\overline{\psi^+_{01}}(v,l)
\label{4.23}
\eeqs
In order to get the characterization equation satisfied by the spectral
data $S_i(k,l)$ in the reduced case $r=\sigma_0\overline q$ we consider
the two Jost solutions
\beqs
\nonumber
{\mbox{\boldmath$\psi$}}^+_1(l)&=&
\left(\bmatrix{c}\psi^+_{01}(l)\\0
\ematrix\right)+G^+_1{\mbox{\boldmath$\psi$}}^+_1(l)\\
{\mbox{\boldmath$\psi$}}^-_2(k)&=&
\left(\bmatrix{c}0\\\psi^-_{02}(k)
\ematrix\right)+G^-_2{\mbox{\boldmath$\psi$}}^-_2(k)
\label{4.24}
\eeqs
and we note that from (\ref{4.1}), for $r=\sigma_0\overline q$, it
easily follows that
\begin{equation}
\label{4.25}
\left(\psi^+_{11}(l)\overline{\psi^{-}_{12}}(k)\right)_u=
\sigma_0\left(\psi^+_{21}(l)\overline{\psi^{-}_{22}}(k)\right)_v.
\end{equation}
By integrating it in the $(u,v)$ plane and by using (\ref{4.24}) we get
\begin{equation}
\label{4.26}
\overline{S_2(k,l)}=\sigma_0S_1(l,k).
\end{equation}
{}From (\ref{4.12}) we derive that the spectral data evolve in time as
follows
\beqs
S_1(k,l,t)&=&e^{i(k^2+l^2)t}S_1(k,l,0)\\
S_2(k,l,t)&=&e^{-i(k^2+l^2)t}S_2(k,l,0)
\eeqs
coherently with the characterization equation (\ref{4.26}).

The discrete component $R_d(k,l)$ of the Spectral Transform has different
possible
characterization. We consider two of them whose matrix elements are
linear combinations of $\delta$ distributions in the complex $k$--plane
and $l$--plane.

The first one is given by the formula
\beqs
\label{4.27}
R_d(k,l)&=&-2\pi i\left(\bmatrix{cc}
R_{11}(k,l)&R_{12}(k,l)\\
R_{21}(k,l)&R_{22}(k,l)
\ematrix\right)\\
R_{11}(k,l)&=&
\sum_{n=1}^N\sum_{n'=1}^N\tau^{nn'}_{11}
\delta(l-\overline{\lambda}_{n'}) \delta(k-\lambda_n)\nonumber\\
R_{12}(k,l)&=&
\sum_{n=1}^N\sum_{m=1}^M\tau^{mn}_{12}
\delta(l-\overline{\lambda}_{n}) \delta(k-\mu_m)\nonumber\\
R_{21}(k,l)&=&
\sum_{n=1}^N\sum_{m=1}^M\tau^{nm}_{21}
\delta(l-\overline{\mu}_{m}) \delta(k-\lambda_n)\nonumber\\
R_{22}(k,l)&=&
\sum_{m=1}^M\sum_{m'=1}^M\tau^{mm'}_{22}
\delta(l-\overline{\mu}_{m'}) \delta(k-\mu_m)\nonumber
\eeqs
where $\lambda_n$ and $\mu_m$, the so called discrete values of the
spectrum, are, respectively, the locations of poles in the complex
$k$--plane of the first and second column of the Jost matrix solution
$\psi$ and the $\tau_{ij}^{mn}$ are some complex constants to be
related to the initial value of $Q$.

The $\delta$ distribution in the complex plane is defined as
\begin{equation}
\label{4.28}
\int\!\!\!\int_\bbbc dl\wedge d\overline{l}\,\delta(l-l_0)f(l)=f(l_0)
\end{equation}
If we introduce the distribution $(l-l_0)\delta(l-l_0)$ operating on
singular functions in the complex plane as follows
\begin{equation}
\label{4.29}
\int\!\!\!\int_\bbbc dl\wedge d\overline{l}\,(l-l_0)\delta(l-l_0)f(l)
=\Res(f,l_0)
\end{equation}
we can deal with a simpler form of the Spectral Transform characterized by
an off
diagonal matrix
\beqs
\label{4.30}
R_d(k,l)&=&-\pi i\left(\bmatrix{cc}
0&-R_{2}(k,l)\\
R_{1}(k,l)&0
\ematrix\right)\\
R_{1}(k,l)&=&
\sum_{n=1}^N\sum_{m=1}^M\left[\tau^{nm}_{1}(l-\mu_m)
\delta(l-\mu_{m}) \delta(k-\lambda_n)+
\widetilde{\tau}_1^{nm} \delta(l-\overline{\mu}_{m})
\delta(k-\lambda_n)\right] \nonumber\\
R_{2}(k,l)&=&
\sum_{n=1}^N\sum_{m=1}^M\left[\tau^{mn}_{2}(l-\lambda_n)
\delta(l-\lambda_{n}) \delta(k-\mu_m)+
\widetilde{\tau}_2^{mn} \delta(l-\overline{\lambda}_{n})
\delta(k-\mu_m)\right]. \nonumber
\eeqs
By inserting the two different characterization of $R_d$ in the
$\overline\partial$--equation (\ref{4.8}) one derives easily that they
are equivalent and, more precisely, that the two set of spectral data
are related by the following equations
\beqs
\tau_{11}&=&-(4+\tau_1\tau_2)^{-1}\tau_1\widetilde{\tau}_2 \nonumber\\
\tau_{12}&=&-(4+\tau_2\tau_1)^{-1}2\widetilde{\tau}_2 \nonumber\\
\tau_{21}&=&(4+\tau_1\tau_2)^{-1}2\widetilde{\tau}_1 \nonumber\\
\tau_{22}&=&-(4+\tau_2\tau_1)^{-1}\tau_2\widetilde{\tau}_1
\label{4.31}
\eeqs
and
\beqs
\tau_1&=&2\tau_{11}\tau_{12}^{-1} \nonumber\\
\tau_2&=&-2\tau_{22}\tau_{21}^{-1} \nonumber\\
\widetilde{\tau}_1&=&2\tau_{21}-2\tau_{11}\tau_{12}^{-1}\tau_{22}
\nonumber\\
\widetilde{\tau}_2&=&-2\tau_{12}+2\tau_{22}\tau_{21}^{-1}\tau_{11}.
\label{4.32}
\eeqs
Without any loss of generality we can consider $M=N$. The cases $M<N$
and $M>N$ are recovered from the special case $M=N$ by choosing some
$\mu$ or $\lambda$ equal.

In order to distinguish between parameters that are fixed by the choice
of the boundaries, i.e. the external data given at all times, and
parameters that are solely connected to the initial data at time $t=0$
it is convenient to parameterize the spectral data in
(\ref{4.30}) as follows
\beqs
\tau_1^{nm}&=&\rho_{nm}
\exp[i\mu_mu_{om}+i\lambda_nv_{on}+i{\mu}^2_mt+i\lambda^2_nt]
\nonumber\\
\widetilde{\tau}_1^{nm}&=&\sum_{p=1}^N\rho_{np}C_{pm}{\mu\cut}_{mm}
\exp[i\overline{\mu}_mu_{om}+i\lambda_nv_{on}+
i\overline{\mu}^2_mt+i\lambda^2_nt] \nonumber\\
\tau_2^{mn}&=&\eta_{mn}
\exp[-i\mu_mu_{om}-i\lambda_nv_{on}-i{\mu}^2_mt-i\lambda^2_nt]
\nonumber\\
\widetilde{\tau}_2^{mn}&=&\sum_{p=1}^N\eta_{mp}D_{pn}{\lambda\cut}_{nn}
\exp[-i{\mu}_mu_{om}-i\overline{\lambda}_nv_{on}-
i{\mu}^2_mt-i\overline{\lambda}^2_nt]
\label{4.33}
\eeqs
where the time dependence is explicitly given,
\beqs
{\mu\cut}_{mn}&=&\overline{\mu}_m-\mu_n\nonumber\\
{\lambda\cut}_{mn}&=&\overline{\lambda}_m-\lambda_n,
\label{4.34}
\eeqs
$\rho$ and $\eta$ are arbitrary complex constant matrices, $v_{on}$ and
$u_{on}$ are arbitrary real constants. We shall show that  the choice
of the boundaries $a^{(1)}_0$ and $a^{(2)}_0$ determines uniquely the
complex matrices $C$ and $D$, together with the
$\lambda$ and the $\mu$, and that the other parameters are left free.
For definiteness we choose
\begin{equation}
\mbox{Im}\,\lambda_n<0,\qquad
\mbox{Im}\,\mu_n>0.  \label{4.35}
\end{equation}
By computing the residua at the
poles $k=\lambda_n$ and $k=\mu_n$ of both sides of the integral
equations (\ref{4.3}) we get
\beqs
\Res(\bpsi_1,\lambda_n)&=&\left(\bmatrix{c}
\Res(\psi_{01},\lambda_n)\\0\ematrix\right)+
G_1^-\Res(\bpsi_1,\lambda_n)\nonumber\\
\Res(\bpsi_2,\mu_n)&=&\left(\bmatrix{c}
0\\\Res(\psi_{02},\mu_n)\ematrix\right)+
G_2^+\Res(\bpsi_2,\mu_n).\label{4.36}
\eeqs
By inserting these integral equations into the r.h.s. of equation
(\ref{4.8}) considered at the special values $k=\lambda_n$ and
$k=\mu_n$ and by recalling that the $\overline\partial$--derivative of
a pole at $k=k_0$ is given by
\begin{equation}
\frac{\partial}{\partial\overline k}\frac{1}{k-k_0}=-2\pi
i\delta(k-k_0)
\label{4.37}
\end{equation}
and the symmetry property of the Green operator $G$
\beqs
G^+_2&=&G^-_1\nonumber\\
G_2^-&=&G_1^-+{1\over2}\left(\bmatrix{cc}
0&\int\!\!du\,q(u,v)\\0&0\ematrix\right)\nonumber\\
G_1^+&=&G_2^++{1\over2}\left(\bmatrix{cc}
0&0\\\int\!\!dv\,r(u,v)&0\ematrix\right)
\label{4.38}
\eeqs
we get, for the special choice of $R_d(k,l)$ in (\ref{4.30}),
\beqs
\Res(\bpsi_1,\lambda_n)&=&\mbox{\bfcal F}_1+
G_1^-\Res(\bpsi_1,\lambda_n)\nonumber\\
\Res(\bpsi_2,\mu_n)&=&\mbox{\bfcal F}_2+
G_2^+\Res(\bpsi_2,\mu_n)\label{4.39}
\eeqs
where
\begin{equation}
\mbox{\bfcal F}_1={{\cal F}_{11}\choose {\cal F}_{21}}
\label{4.40}
\end{equation}
with
\beqs
{\cal F}_{11}&= &{1\over4}\sum_{m=1}^N
\!\!\int\!\!du\,q\psi_{22}^-(\overline{\mu}_{m}) (\rho
C)_{nm}{\mu\cut}_{mm}
\exp[i\overline{\mu}_mu_{om}+i\lambda_nv_{on}+
i\overline{\mu}^2_mt+i\lambda_n^2t]\nonumber\\
{\cal F}_{21}&=
&{1\over2}\sum_{m=1}^N\rho_{nm}\left\{\Res(\psi_{02}^+,\mu_m)
\exp[i{\mu}_mu_{om}+i\lambda_nv_{on}+
i{\mu}^2_mt+i\lambda_n^2t]\right.+\nonumber\\
&&\sum_{p=1}^N\left.C_{mp}{\mu\cut}_{pp}\psi_{02}^-(\overline{\mu}_p)
\exp[i\overline{\mu}_pu_{op}+i\lambda_nv_{on}+
i\overline{\mu}^2_pt+i\lambda_n^2t]\right\}
\label{4.41}
\eeqs
and
\begin{equation}
\mbox{\bfcal F}_2={{\cal F}_{12}\choose {\cal F}_{22}}
\label{4.42}
\end{equation}
with
\beqs
{\cal F}_{22}&= &-{1\over4}\sum_{m=1}^N
\!\!\int\!\!dv\,r\psi_{11}^+(\overline{\lambda}_{m}) (\eta
D)_{nm}{\lambda\cut}_{mm}
\exp[-i{\mu}_nu_{on}-i\overline{\lambda}_mv_{om}-
i{\mu}^2_nt-i\overline{\lambda}_m^2t]\nonumber\\
{\cal F}_{12}&=
&-{1\over2}\sum_{m=1}^N\eta_{nm}\left\{\Res(\psi_{01}^-,\lambda_m)
\exp[-i{\mu}_nu_{on}-i\lambda_mv_{om}-
i{\mu}^2_nt-i\lambda_m^2t]\right.+\nonumber\\
&&\sum_{p=1}^N\left.
D_{mp}{\lambda\cut}_{pp}\psi_{01}^+(\overline{\lambda}_p)
\exp[-i{\mu}_nu_{on}-i\overline{\lambda}_pv_{op}-
i{\mu}^2_nt-i\overline{\lambda}_p^2t]\right\}
\label{4.43}
\eeqs
By comparing the integral equations (\ref{4.36}) and (\ref{4.39}) one
gets
\beqs
\lefteqn{\Res(\psi_{01}^-,\lambda_n)=}&&\nonumber\\
&&{1\over4}\sum_{m=1}^N
\!\!\int\!\!du\,q\psi_{22}^-(\overline{\mu}_{m}) (\rho
C)_{nm}{\mu\cut}_{mm}
\exp[i\overline{\mu}_mu_{om}+i\lambda_nv_{on}+
i\overline{\mu}^2_mt+i\lambda_n^2t]\label{4.44}\\
\lefteqn{\Res(\psi_{02}^+,\mu_n)=}&&\nonumber\\
&&-{1\over4}\sum_{m=1}^N
\!\!\int\!\!dv\,r\psi_{11}^+(\overline{\lambda}_{m}) (\eta
D)_{nm}{\lambda\cut}_{mm}
\exp[-i{\mu}_nu_{on}-i\overline{\lambda}_mv_{om}-
i{\mu}^2_nt-i\overline{\lambda}_m^2t]\qquad\qquad
\label{4.45}
\eeqs
and
\beqs
&&\Res(\psi_{01}^-,\lambda_n)=-
\sum_{m=1}^N D_{nm}{\lambda\cut}_{mm}\psi_{01}^+(\overline{\lambda}_m)
\exp[i\lambda_nv_{on}-i\overline{\lambda}_mv_{om}
+i\lambda_n^2t-i\overline{\lambda}^2_mt]\label{4.46}
\\
&&\Res(\psi_{02}^+,\mu_n)=-
\sum_{m=1}^N C_{nm}{\mu\cut}_{mm}\psi_{02}^-(\overline{\mu}_m)
\exp[-i\mu_nu_{on}+i\overline{\mu}_mu_{om}
-i\mu_n^2t+i\overline{\mu}^2_mt]\qquad\qquad
\label{4.47}
\eeqs
Equations (\ref{4.46}) and (\ref{4.47}) determine uniquely the boundary
values $a^{(1)}_0$ and $a^{(2)}_0$ of the auxiliary field
$A=\diag(A^{(1)},A^{(2)})$. In fact, they determine the Spectral
Transform of $a^{(1)}_0$ and $a^{(2)}_0$ considered as potentials in
the time dependent Schr\"odinger equations of (\ref{4.13}). If we use
the Spectral Transform defined in the previous section
\beqs
\frac{\partial\psi_{01}}{\partial\overline k}&=&
\!\!\int\!\!\!\int\!\!dl\wedge d\overline{l}\,
\psi_{01}(l)r_1(k,l)e^{i(k^2-l^2)t}\label{4.48}\\
\frac{\partial\psi_{02}}{\partial\overline k}&=&
\!\!\int\!\!\!\int\!\!dl\wedge d\overline{l}\,
\psi_{02}(l)r_2(k,l)e^{-i(k^2-l^2)t}
\label{4.49}
\eeqs
we get that
\beqs
\label{4.50}
r_1(k,l)&=&2\pi i\sum_{n,m=1}^N\exp[i\lambda_nv_{on}-
i\overline{\lambda}_mv_{om}]D_{nm}{\lambda\cut}_{mm}
\delta(l-\overline{\lambda}_m)\delta(k-\lambda_n)\\
r_2(k,l)&=&2\pi i\sum_{n,m=1}^N\exp[-i\mu_nu_{on}+
i\overline{\mu}_mu_{om}]C_{nm}{\mu\cut}_{mm}
\delta(l-\overline{\mu}_m)\delta(k-\mu_n)
\label{4.51}
\eeqs
It results, in particular, that  $\psi_{01}$ and $\psi_{02}$ have
simple poles at $k=\lambda_n$ and at $k=\mu_n$ as expected.

If we consider the KPI equation associated, for instance, to the time
dependent Schr\"odinger equation  for $a^{(1)}_0(v,t)$ by taking into
account that $(v,t)$ are to be considered as ``space'' variables of the
KPI equation while its ``time'' variable has to be considered as an
additional parameter of the potential $a^{(1)}_0(v,t)$, it results
that the obtained boundary value $a^{(1)}_0(v,t)$ coincides with an $N$
wave soliton solution of the KPI equation in the $(v,t)$ plane at some
fixed ``time''. Analogously, $a^{(2)}_0(u,t)$ can be considered as an N
wave soliton solution of a KPI equation in the $(u,t)$ ``plane'' at
some fixed ``time''. The real parameters $v_{on}$ and $u_{on}$ fix the
position of the $\mbox{n}^{\mbox{th}}$ wave soliton in the
corresponding ``plane''.

If we require $a^{(1)}_0$ and $a^{(2)}_0$ to be real the complex
matrices $C$ and $D$ satisfy
\beqs
D_{nm}{\lambda\cut}_{mm}&=&\overline{D}_{mn}{\lambda\cut}_{nn}
\label{4.52}\\
C_{nm}{\mu\cut}_{mm}&=&\overline{C}_{mn}{\mu\cut}_{nn}.
\label{4.53}
\eeqs
Moreover, without any loss of generality, by eventually shifting
$u_{on}$ and $v_{on}$ we can choose
\begin{equation}
D^2_{nn}=C^2_{nn}=1
\label{4.54}
\end{equation}
Equations (\ref{4.44}) and (\ref{4.45}) solve the direct problem
furnishing the matrices $\rho$ and $\eta$ in terms of the
eigenfunction of the spectral problem (\ref{4.1}), of $Q$ and of the
parameters and eigenfunctions defining the boundary. By using the
orthogonality relations derived in the previous section
\beqs
&&i\!\!\int\!\!dv\,\psi_{01}^+(\overline{\lambda}_{m})
\overline{\Res(\psi_{01}^-,\lambda_n)}=\delta_{mn}\nonumber\\
&&i\!\!\int\!\!du\,\psi_{02}^-(\overline{\mu}_{m})
\overline{\Res(\psi_{02}^+,\mu_n)}=\delta_{mn}\label{4.55}
\eeqs
we obtain the formulae
\beqs
&&\rho_{nm}=-i\exp[-i\lambda_nv_{on}-i\lambda_n^2t]
{\left(X^{-1}\right)}_{nm}\nonumber\\
&&\eta_{nm}=-i\exp[i\mu_nu_{on}+i\mu^2_nt] {\left(Y^{-1}\right)}_{nm}
\eeqs
where
\beqs
&&X_{ns}={1\over4}\sum_{m=1}^NC_{nm}
{\mu\cut}_{mm}e^{i\overline{\mu}_mu_{om}+i \overline{\mu}^2_mt}
\!\!\int\!\!\!\int\!\!du\,dv\,
q\psi_{22}^-(\overline{\mu}_m)
\overline{\psi_{01}^+}(\overline{\lambda}_s)\nonumber\\
&&Y_{ns}={1\over4}\sum_{m=1}^ND_{nm}
{\lambda\cut}_{mm}e^{-i\overline{\lambda}_mv_{om}-i
\overline{\lambda}^2_mt} \!\!\int\!\!\!\int\!\!du\,dv\,
r\psi_{11}^+(\overline{\lambda}_m)
\overline{\psi_{02}^-}(\overline{\mu}_s).
\label{4.57}
\eeqs
It is worth noting that the chosen parameterization of the spectral data
in (\ref{4.33}) allows us to discriminate between the parameters
$\lambda_n$, $\mu_n$, $u_{on}$, $v_{on}$, $C_{mn}$, $D_{mn}$ which are
fixed by the choice of the boundary values $a^{(1)}_0$ and $a^{(2)}_0$
and parameters $\rho_{mn}$, $\eta_{mn}$ which are left free and are
expected to govern the specific nonlinear dynamic of the DSI equation.
In analogy with the one dimensional case we call $\rho_{mn}$ and
$\eta_{mn}$ the normalization matrix coefficients.

In the reduced case $r=\sigma_0\overline q$ from equation (\ref{4.26})
computed at $k=\mu_n$ and at $l=\lambda_m$, by using the orthogonality
relations (\ref{4.55}), we get the necessary condition
\begin{equation}
\sum_{m=1}^N\rho_{nm}C_{ms}{\mu\cut}_{ss}=
\sigma_0\sum_{m=1}^ND_{nm}{\lambda\cut}_{mm}\overline{\eta}_{sm}
\label{4.58}
\end{equation}
In solving the inverse problem we shall prove that this condition is
also sufficient for having $r=\sigma_0\overline q$.

We are left with the solution of the inverse problem, i.e. we have
to reduce the reconstruction of the matrix $Q$ and the auxiliary field
$A$ from given spectral data $R(k,l)$ to the solution of a linear
problem.

The $\overline\partial$--equation (\ref{4.8}) together with the
asymptotic requirement in (\ref{4.2}) defines a non--local
Riemann--Hilbert problem for the matrix function
\begin{equation}
\phi(x,y,t,k)=\psi(x,y,t,k)e^{-ik(\sigma_3x-y)}
\label{4.59}
\end{equation}
Its solution is obtained by solving the singular linear integral
equation (the space--time variables of $\phi$ are understood)
\begin{equation}
\phi(k)=\bbbone +{1\over2\pi i}\!\!\int\!\!\!\int_{\bbbc}\!\!
\frac{dh\wedge d\overline{h}}{h-k}
\!\!\int\!\!\!\int_{\bbbc}\!\!dl\wedge d\overline{l}\,\phi(l)
e^{il(\sigma_3x-y)}R(h,l,t)e^{-ih(\sigma_3x-y)}
\label{4.60}
\end{equation}
This equation furnishes an asymptotic expansion in powers of $1\over k$
of $\phi$
\begin{equation}
\phi=\bbbone +{1\over k}\phi^{(1)}+O\left({1\over k^2}\right)
\label{4.61}
\end{equation}
where
\begin{equation}
\phi^{(1)}=-{1\over2\pi i}\!\!\int\!\!\!\int_{\bbbc}\!\!
dh\wedge d\overline{h}
\!\!\int\!\!\!\int_{\bbbc}\!\!dl\wedge d\overline{l}\,\phi(l)
e^{il(\sigma_3x-y)}R(h,l,t)e^{-ih(\sigma_3x-y)}.
\label{4.62}
\end{equation}
Then the solution of the inverse problem is achieved by inserting the
expansion into the Zakharov--Shabat spectral problem $T_1\psi=0$ and into the
auxiliary spectral problem $T_2\psi=\psi\Omega$ and by identifying the
coefficients of the powers of $1\over k$. One obtains $Q$, $A$ and a
useful expression for $Q^2$
\beqs
&&Q=i[\sigma_3,\phi^{(1)}]\label{4.63}\\
&&A=-i(\partial_x-\sigma_3\partial_y)\diag\phi^{(1)}\label{4.64}\\
&&Q^2=2i\sigma_3(\partial_x+\sigma_3\partial_y)\diag\phi^{(1)}.
\label{4.65}
\eeqs
When only the discrete part of the spectrum is present one can derive
explicit algebraic formulae for $Q$ and $A$. The requirement that
$\phi$ has simple poles at $k=\lambda_n$ and at $k=\mu_n$ and that
$\phi$ goes to $\bbbone$ in the large $k$ limit fixes a $k$ dependence
of the form
\begin{equation}
\phi=\bbbone
+\sum_{n=1}^N\left(\frac{\bvarphi_{n1}e^{\beta_n}}{k-\lambda_n},
\frac{\bvarphi_{n2}e^{\alpha_n}}{k-\mu_n}\right)
\label{4.66}
\end{equation}
where, for convenience, in the residua an explicit exponential factor
has been introduced with
\beqs
\alpha_n&=&i\mu_n(u-u_{on})-i\mu^2_nt \label{4.67}\\
\beta_n&=&-i\lambda_n(v-v_{on})+i\lambda^2_nt
\label{4.68}
\eeqs
The vectors $\bvarphi_{n1}$ and $\bvarphi_{n2}$ are computed by
inserting (\ref{4.66}) into the $\overline\partial$--equation
(\ref{4.8}) with $R(k,l)$ from (\ref{4.30}) and (\ref{4.33}). One
obtains
\beqs
\left[\bbbone+{1\over4}\eta{\cal B}\rho{\cal A}\right] \bvarphi_2&=&
-{1\over2}\eta\delta{1\choose 0}-{1\over4}\eta{\cal B}\rho\gamma
{0\choose1}\label{4.69}\\
\left[\bbbone+{1\over4}\rho{\cal A}\eta{\cal B}\right] \bvarphi_1&=&
{1\over2}\rho\gamma{0\choose1}-{1\over4}\rho{\cal A}\eta\delta
{1\choose 0}\label{4.70}
\eeqs
where
\beqs
&&\gamma_n=\sum_{m=1}^NC_{nm}{\mu\cut}_{mm}e^{\overline{\alpha}_m}
\label{4.71}\\
&&\delta_n=\sum_{m=1}^ND_{nm}{\lambda\cut}_{mm} e^{\overline{\beta}_m}
\label{4.72}\\
&&{\cal A}=\bbbone+C\alpha,\qquad \alpha_{nm}=
\frac{{\mu\cut}_{nn}}{{\mu\cut}_{nm}}e^{\overline{\alpha}_n+\alpha_m}
\label{4.73}\\
&&{\cal B}=\bbbone+D\beta,\qquad \beta_{nm}=
\frac{{\lambda\cut}_{nn}}{{\lambda\cut}_{nm}}
e^{\overline{\beta}_n+\beta_m}. \label{4.74}
\eeqs
{}From equation (\ref{4.63}) one gets
\beqs
Q=2i\sigma_3\sum_{n=1}^N\left(\bmatrix{cc}
0&\left(\bvarphi_1\right)_2e^{\alpha_n}\\
\left(\bvarphi_2\right)_1e^{\beta_n}&0
\ematrix\right)\label{4.75}
\eeqs
and from (\ref{4.64}) and (\ref{4.65}) by the rule for differentiating
a determinant one gets
\beqs
&&A=2\left(\bmatrix{cc}
\partial^2_v&0\\
0&-\partial_u^2\ematrix\right)\ln\Delta\label{4.76}\\
&&Q^2=-4\partial_u\partial_v\ln\Delta
\label{4.77}
\eeqs
where
\begin{equation}
\Delta=\det\left[\bbbone+{1\over4}\eta{\cal B}\rho{\cal A}\right] =
\det\left[\bbbone+{1\over4}\rho{\cal A}\eta{\cal B}\right].
\label{4.78}
\end{equation}
By solving. in a similar way, the $\overline\partial$--equations
(\ref{4.48}) and (\ref{4.49}) one can derive also the boundaries
\beqs
a^{(1)}_0&=&2\partial^2_v\ln\det{\cal B}\label{4.79}\\
a^{(2)}_0&=&-2\partial^2_u\ln\det{\cal A}.
\label{4.80}
\eeqs
In order to study the reduction and the regularity properties of the
found solution it is convenient to introduce the hermitian matrices
\beqs
\widetilde{C}_{mn}&=&iC_{mn}{\mu\cut}_{nn}\label{4.81}\\
\widetilde{D}_{mn}&=&-iD_{mn}{\lambda\cut}_{nn}.
\label{4.82}
\eeqs
One can easily get that
\beqs
\widetilde{C}{\cal A}^{\dag}&=&{\cal A}\widetilde{C}\label{4.83}\\
\widetilde{D}{\cal B}^{\dag}&=&{\cal B}\widetilde{D}\label{4.83'}
\eeqs
Moreover, if the constraint (\ref{4.58}), which rewritten by using
$\widetilde C$ and $\widetilde D$ reads
\begin{equation}
\widetilde{D}\eta^{\dag}=-\sigma_0\rho\widetilde{C},
\label{4.84}
\end{equation}
is satisfied one can easily verify directly that the reduction
condition
\begin{equation}
r=\sigma_0\overline{q}
\label{4.85}
\end{equation}
is fulfilled. We conclude that the constraint (\ref{4.58}) is necessary
and sufficient for having the reduction.

To see that (\ref{4.69}) and (\ref{4.70}) can be solved for
$\bvarphi_1$ and $\bvarphi_2$ and that, consequently, $Q$ and $A$ are
regular, we first note that the two hermitian matrices
\beqs
\widetilde{\alpha}_{nm}&=&-
i\frac{e^{\overline{\alpha}_n+\alpha_m}}{{\mu\cut}_{nm}}
\label{4.86}\\
\widetilde{\beta}_{nm}&=&
i\frac{e^{\overline{\beta}_n+\beta_m}}{{\lambda\cut}_{nm}}
\label{4.87}
\eeqs
are positive definite. The corresponding quadratic forms in the dummy
vector $P$ can be written as
\beqs
&&\sum_{n,m=1}^N\overline{P}_nP_m\widetilde{\alpha}_{nm}=
\sum_{n,m=1}^N\overline{P}_nP_m\!\!\int_u^{+\infty}\!\!
du'\,e^{\overline{\alpha}_n+\alpha_m}
=\!\!\int^{+\infty}_u\!\!du'{\left|\sum_{n=1}^NP_ne^{\alpha_n}\right|}^2
\label{4.88}\\
&&\sum_{n,m=1}^N\overline{P}_nP_m\widetilde{\beta}_{nm}=
\sum_{n,m=1}^N\overline{P}_nP_m\!\!\int_{-\infty}^v\!\!
dv'\,e^{\overline{\beta}_n+\beta_m}
=\!\!\int_{-\infty}^v\!\!dv'{\left|\sum_{n=1}^NP_ne^{\beta_n}\right|}^2
\label{4.89}
\eeqs
which are positive unless all $P_n$ are zero. Therefore if the
matrices $\widetilde C$ and $\widetilde D$ are chosen to be positive
definite, because the product of two positive hermitian matrices is
positive ${\cal A}=\bbbone+C\alpha=
\bbbone+\widetilde{C}\widetilde{\alpha}$ and  ${\cal B}=
\bbbone+D\beta= \bbbone+\widetilde{D}\widetilde{\beta}$  are positive
definite and the boundaries $a^{(1)}_0$ and $a^{(2)}_0$ are regular.
If, moreover, the reduction condition (\ref{4.81}) is satisfied the
matrices (\ref{4.69}) and (\ref{4.70}) can be rewritten as
\beqs
&&\bbbone+{1\over4}\eta{\cal B}\rho{\cal A} =\bbbone -{1\over4}\sigma_0
\eta{\cal B}\widetilde{D}\eta^{\dag}\widetilde{C}^{-1}{\cal A}
\label{4.90}\\
&&\bbbone+{1\over4}\rho{\cal A}\eta{\cal B} =\bbbone -{1\over4}\sigma_0
\rho{\cal A}\widetilde{C}\rho^{\dag}\widetilde{D}^{-1}{\cal B}
\label{4.91}
\eeqs
which for $\sigma_0=-1$ are positive definite for regular $\eta$ and
$\rho$. For $\sigma_0=1$ they are positive definite for $\eta$ and
$\rho$ matrices with sufficiently small norms.

\subsection{The dynamics of localized solitons in two dimensions}
The solution $q$, in the generic case, describes $N^2$ localized
coherent structures interacting in a complicated way at finite times,
but moving in the far past and in the far future with constant
velocities ${\bf V}_{nm}=(2\mbox{Re}\,\lambda_n,2\mbox{Re}\,\mu_m)$ and
without changing form. It is therefore natural to call this solution
the $N^2$ soliton solution. It is parameterized by a point in a space
of $4N(N+1)$ real parameters. Of these parameters $2N(N+2)$ and,
precisely, $\lambda_n$, $\mu_n$, $v_{on}$, $u_{on}$, $C$ and $D$ are
determined by the choice of the boundaries and fix the velocity and the
possible location of the solitons in the plane, while the remaining
$2N^2$ and, precisely, $\eta$ and $\rho$ govern the dynamics of the
solitons during the interaction.

A relevant information in the global dynamical behaviour of the
solitons is furnished by the mass (energy, charge, or number of
particles according to the physical context) of the solution $q$
\begin{equation}
M=\!\!\int\!\!\!\int\!\!|q|^2du\,dv
\label{4.92}
\end{equation}
and by the masses of the solitons at $t=\pm\infty$
\begin{equation}
M^{(\pm )}_{mn}=\!\! \int\!\!\!\int\!\!\mid q^{(\pm )}_{mn}\mid
^{2} du dv\qquad (m,n=1, 2, \ldots ,N) \label{50}
\end{equation}
where
$q^{(\pm )}_{mn}$ is the asymptotic behaviour of $q$ at $t=\pm \infty
$ computed  in the rest reference frame  of  the $(m,n)$  soliton.  If
$\det (\rho \eta )\neq 0, \det \alpha \neq 0$ and $\det  \beta \neq 0$
the mass $M$  is given  by  the  simple  formula
\begin{equation} M =-4 \sigma_0
\ln\frac{\det\left(\bbbone+{1\over4}\eta\rho\right)}
{\det\left({1\over4}\eta\rho\right)}.
\end{equation}
In general, it results that
$M = \sum^{}_{mn}M^{(-)}_{mn}= \sum^{}_{mn}M^{(+)}_{mn}$.
However,  the
mass of the single soliton is not conserved and in  particular  it
can be zero at $t=+\infty $ or at $t=-\infty $.  For  a  special  choice  of
the
parameters the mass of the $(m,n)$ soliton can be zero at $t=\pm
\infty $  also when the coefficients $\rho _{mn}$ and $\eta _{nm}$
are not both equal to  zero.  We call these solitons with  zero  mass
virtual  solitons  and  they generate peculiar effects as in figure 1 of
\cite{dynamics}
where  a  virtual  soliton collides with a soliton forcing it to change
velocity.

On the other hand, the total momentum of $q$
\begin{equation}
{\bf P}
= ( P_{1}, P_{2}),\quad P_{1}= i\!\!\int\!\!\! \int
(\overline{q}_{u}q-\overline{q}q_{u})du\, dv,\quad
P_{2}= i\!\!\int\!\!\! \int (\overline{q}_{v}q-\overline{q}q_{v})du\, dv
\end{equation}
is not conserved and, in fact, it results that
\begin{equation}
{d{\bf P}\over dt} =
\left(2\int\!\!\!\int \mid q\mid ^{2}{\partial \over \partial u}
A_{2}(u,t)du\,dv,-2\int\!\!\!\int \mid q\mid ^{2}
{\partial \over \partial v}A_{1}(v,t)du\,dv\right)
\end{equation}
where
\begin{equation}
A_{1}=a^{(1)}_0- {1\over 4}\sigma_0 \int du{\left(\mid q\mid
^{2}\right)}_{v} , \qquad A_{2}=a^{(2)}_0+ {1\over 4}\sigma_0 \int
dv{\left(\mid q\mid ^{2}\right)}_{u}.  \end{equation}

Because the four--soliton solution displays all  the  richness
of the general case  we  examine  it  in  detail.  We  choose  for
definiteness $\lambda _{n\Im}<0$, $\mu _{n\Im}> 0$ $(n = 1, 2)$,
$(\lambda _{2\Re}-\lambda _{1\Re}) > 0$   and
$(\mu _{2\Re}-\mu _{1\Re}) <0$. It is convenient to parameterize the
matrices $\rho $ and
$\eta $ in such a way that the reduction condition (\ref{4.58}) is
automatically satisfied, i.e.
\beqs
\eta&=&{1\over
d}\left(\bmatrix{cc} (\overline{\ell }_{1}-\overline{\cal
D}m_{1}) \sqrt{\displaystyle \frac{{\mu\cut}_{11}}{{\lambda\cut}_{11}}}&
(m_{1}-{\cal
D}\overline{\ell }_{1})
\sqrt{\displaystyle \frac{{\mu\cut}_{11}}{{\lambda\cut}_{22}}}\\[10pt]
(\overline{m}_{2}-\overline{{\cal D}}\ell _{2})
\sqrt{\displaystyle \frac{{\mu\cut}_{22}}{{\lambda\cut}_{11}}}&
(\ell _{2}-{\cal D}\overline{m}_{2})
\sqrt{\displaystyle \frac{{\mu\cut}_{22}}{{\lambda\cut}_{22}}}
\ematrix\right)
\label{51}\\
\rho & =&-{\sigma_0\over c}\left(\bmatrix{cc}
(\ell_{1}-\overline{\cal C}m_{2})
\sqrt{\displaystyle \frac{{\lambda\cut}_{11}}{{\mu\cut}_{11}}}&
(m_{2}-{\cal C}\ell_{1})
\sqrt{\displaystyle \frac{{\lambda\cut}_{11}}{{\mu\cut}_{22}}}\\[10pt]
(\overline{m}_{1}-\overline{{\cal C}}\overline{\ell}_{2})
\sqrt{\displaystyle \frac{{\lambda\cut}_{22}}{{\mu\cut}_{11}}}&
(\overline{\ell}_{2}-{\cal C}\overline{m}_{1})
\sqrt{\displaystyle \frac{{\lambda\cut}_{22}}{{\mu\cut}_{22}}}
\ematrix\right)
\label{52}
\eeqs
with $\ell _{n}$ and $m_{n}$ $(n=1,2)$
arbitrary complex parameters and
\begin{equation}
{\cal C} = C_{12}\sqrt{\frac{{\mu\cut}_{22}}{{\mu\cut}_{11}}},
\qquad {\cal D} =
D_{12}\sqrt{\frac{{\lambda\cut}_{22}}{{\lambda\cut}_{11}}}
\label{53}
\end{equation}
\begin{equation}
c = 1 - \mid {\cal C}\mid ^{2},\qquad d = 1 -\mid {\cal D}\mid ^{2}.
\label{54}
\end{equation}
The parameters $\lambda _{n}, \mu _{n}, {\cal C}$
and ${\cal D}$ fix the boundaries while $\ell _{n}$ and $m_{n}$
govern the dynamics of the soliton interaction. In the case $\sigma_0=-
1$ if $c > 0$ and $d > 0$ the solution is regular.

The masses  of
the  solitons  can be  explicitly   computed.  For a generic choice of
the parameters they do not depend  on  the spectral parameters
$\lambda_{m}$,  $\mu _{n}$ and on the initial positions
$(u_{om},v_{on})$.  We get $(\sigma_0=-1)$
\beqs
M^{(-)}_{11}&=&4   \ln \left(1 + 4
\frac{c\mid\ell_2-{\cal D}\overline m_2\mid^2}
{\mid\overline{\ell}_{1}\ell_{2}-m_{1}\overline{m}_{2}\mid^{2}
+ 4   cd \mid m_{2}\mid
^{2}}\right)\\ M^{(-)}_{22}&=&4   \ln \left(1 + 4
\frac{d\mid\ell_1-\overline{\cal C}m_2\mid^2} {\mid \overline{\ell
}_{1}\ell _{2}- m_{1}\overline{m}_{2}\mid ^{2} + 4   cd \mid m_{2}\mid
^{2}}\right)\\ M^{(+)}_{11}&=&4   \ln \left(1 + 4
\frac{d\mid\ell_2-\overline{\cal C}m_1\mid^2} {\mid \overline{\ell
}_{1}\ell _{2}- m_{1}\overline{m}_{2}\mid ^{2} + 4   cd \mid m_{1}\mid
^{2}}\right)\\ M^{(+)}_{22}&=&4   \ln \left(1 + 4
\frac{c\mid\ell_1-{\cal D}\overline m_1\mid^2} {\mid \overline{\ell
}_{1}\ell _{2}- m_{1}\overline{m}_{2}\mid ^{2} + 4   cd \mid m_{1}\mid
^{2}}\right)\\
M^{(-)}_{12}&=&4\ln\bigg(1+4
\frac{\mid 4dc\overline  m_2+(\ell_2\overline\ell_1
-\overline m_2 m_1)[(\ell_1-\overline{\cal C} m_2)\overline{\cal D}-
(\overline m_1-\overline{\cal C}\overline\ell_2)]\mid^2}
{[\mid\overline{\ell}_{1}\ell_{2}-m_{1}\overline{m}_{2}\mid^2
+4c(\mid m_2-\overline\ell_2
{\cal D}\mid^2+d\mid\ell_2\mid^2)]}\times
\nonumber\\
\lefteqn{\frac{1}
{[\mid\overline{\ell}_{1}\ell_{2}-m_{1}\overline{m}_{2}\mid^2+
4d(\mid m_2-\ell_1{\cal C}\mid^2+c\mid\ell_1\mid^2)]}
\bigg)}&&\\
M^{(-)}_{21}&=&4\ln\left(1+4dc\frac{\mid m_2\mid^2}
{\mid\overline{\ell}_{1}\ell_{2}-m_{1}\overline{m}_{2}\mid^2}\right)\\
M^{(+)}_{12}&=&4\ln\left(1+4dc\frac{\mid m_1\mid^2}
{\mid\overline{\ell}_{1}\ell_{2}-m_{1}\overline{m}_{2}\mid^2}\right)\\
M^{(+)}_{21}&=&4\ln\bigg(1+4
\frac{\mid 4dc  m_1+(\ell_2\overline\ell_1-\overline m_2 m_1)
[(\overline\ell_2-{\cal C}\overline m_1){\cal D}-
( m_2-{\cal C}\ell_1)]\mid^2}
{[\mid\overline{\ell}_{1}\ell_{2}-m_{1}\overline{m}_{2}\mid^2
+4c(\mid m_1-\overline\ell_1
{\cal D}\mid^2+d\mid\ell_1\mid^2)]}\times\nonumber\\
\lefteqn{\frac{1}{[\mid\overline{\ell}_{1}\ell_{2}-m_{1}
\overline{m}_{2}\mid^2+
4d(\mid m_1-\ell_2{\cal C}\mid^2+c\mid\ell_2\mid^2)]}\bigg)}&&
\eeqs
Of special interest are the cases in which one or more masses
are zero. For  definiteness,  we  choose  the  masses
$M^{(\pm )}_{mn}$ $(m\neq n)$
different from zero  and  we  consider  the  case  in  which  some
$M^{(\pm )}_{mm}$ are zero.

For $\ell _{1}= \overline{{\cal C}}m_{2}$ or
$\ell _{2}= {\cal D}\overline{m}_{2}$ one gets
$M^{(-)}_{22}= 0$ or $M^{(-)}_{11}= 0$  and  the
solution describes the creation  of  a  soliton.
For $\ell _{1}= {\cal D}\overline{m}_{1}$  or
$\ell _{2}= \overline{{\cal C}}m_{1}$ one gets
$M^{(+)}_{22}= 0$ or $M^{(+)}_{11}= 0,$ i.e. the  annihilation  of  a
soliton. For $\ell _{1}= \overline{{\cal C}}m_{2}$
and $\ell _{2}= {\cal D}\overline{m}_{2}$ it results that
$M^{(-)}_{22}= M^{(-)}_{11}= 0$ and
the solution describes the creation of a  pair  of  solitons.  For
$\ell _{1}= {\cal D}\overline{m}_{1}$ and
$\ell _{2}= \overline{{\cal C}}m_{1}$ one gets
$M^{(+)}_{22}= M^{(+)}_{11}= 0$ , i.e. the annihilation
of  a  pair  of  solitons.
For $\ell _{1}= \overline{{\cal C}}m_{2}$  and
$\ell _{2}= \overline{{\cal C}}m_{1}$  one   gets
$M^{(-)}_{22}= M^{(+)}_{11}= 0,$ i.e. a soliton changes its
mass and velocity.  For
$\ell _{1}= {\cal D}\overline{m}_{1}$,
$\ell _{2}=(\mid {\cal D}\mid ^{2}/{\cal C})m_{1}$,
$m_{2}=({\cal D}/\overline{{\cal C}})\overline{m}_{1}$ one gets
$M^{(\pm )}_{22}= M^{(-)}_{11}= 0$ and the
solution  describes  the  creation  of  a  soliton.
For $\ell _{1}= {\cal D}\overline{m}_{1}$,
$\ell _{2}=(\mid {\cal C}\mid ^{2}/\overline{{\cal D}})\overline{m}_{2}$,
$m_{2}=({\cal D}/\overline{{\cal C}})\overline{m}_{1}$
one gets $M^{(\pm )}_{22}= M^{(+)}_{11}= 0$ and the solution
describes  the  annihilation  of  a   soliton.
For $\ell _{1}= \ell _{2}= \overline{{\cal C}}m$,
$m_{1}= m_{2}= m$, $\overline{{\cal C}}m = {\cal D}\overline{m}$
all the masses $M^{(\pm )}_{mm}$ are  zero  and  one  gets  a
solution describing two interacting solitons.

All these dynamical behaviours can be  obtained  by  choosing
the same boundaries. Boundaries do not give any information on the
dynamics of solitons, but fix only their  kinematics,  i.e.  their
possible locations in the plane and their velocities.

Of special interest is the case in which  two  spectral  data
are equal, say $\mu _{1}= \mu _{2}$. In general,  this  solution  describes
the
interaction of two solitons.  The  masses $M^{(\pm )}_{mm} (m=1,2)$
in  this degenerate case  depend  also  on  the  initial  position  of
the solitons.  With a special choice of the  parameters
one can get a solution describing the fission of a  soliton  (see
figure 2 in \cite{dynamics}) and the fusion of two solitons. If in addition
one chooses
$\det (\rho \eta ) = 0$ the solution describes a single soliton  that  by  the
interaction with a virtual soliton is forced  to  change  velocity
(see figure 1 in \cite{dynamics}). Note that the two dynamical  processes
described  in figure 1 and figure 2 in \cite{dynamics} are obtained by
choosing the same boundaries  and different matrices $\rho $ and $\eta $.

\subsection{Asymptotic bifurcation of multidimensional solitons}
The degeneracy of the solution when two discrete spectral data are
chosen to be equal is worth of a deeper analysis.

This phenomenon is well known in one dimension, but, while in one
dimension by taking two eigenvalues equal in the $N$--soliton solution
we recover the $(N{-}1)$--soliton solution, in 2+1 dimensions, as shown
by the previous example, we get a new solution.

To understand the underlaying mechanism it is convenient to consider
the simplest case in which the effect takes place. Precisely,
we choose $N=2$, the matrices $\rho$ and $\eta$ diagonal
\begin{equation}
\rho=\diag(\rho_1,\rho_2),\qquad \eta=\diag(\eta_1,\eta_2)
\end{equation}
and
\begin{equation}
D=\bbbone,\qquad C=\bbbone.
\end{equation}
In the reduced case $\sigma_0=-1$ (we are considering) the two matrices
$\eta$ and $\rho$ are related by the constraint
\begin{equation}
\rho_n{\mu\cut}_{nn}=-\overline{\eta}_n{\lambda\cut}_{nn}.
\end{equation}
The set of these solutions describes a family of geometrical objects
evolving in time in the $(x,y)$ plane; each object $E_{(p,t)}$ of the
family is parameterized by a point $p$ in a $P$ space of 16 real
parameters and by the time and therefore corresponds to a point in a
17--dimensional space $S=\{p,t\}$. We call generically stable or
generic those objects $E_{s_0}$ of the family that depend in a
differentiable way on the 17 parameters in a neighborhood of a point
$s_0$. According to the usual definition in catastrophe theory we call
the complement of this open set $\{s_0\}$ the set of bifurcation
points.

The generic solution describes two solitons mutually interacting
without changing shape and velocity. The only effect of the interaction
is a shift in the position and in the overall phase.

The only bifurcation points are $(p,-\infty)$ and $(p,+\infty)$ with a
special choice of the parameters $p$. Precisely,
when any couple of discrete eigenvalues $\lambda_n$, $\lambda_m$ or
$\mu_n$, $\mu_m$ have the same real part, i.e. when the parameters
belong to the hyperplanes $\lambda_{n\Re}=\lambda_{m\Re}$ or
$\mu_{n\Re}=\mu_{m\Re}$ in $P$, the two--soliton solution is not stable
at large times.  For this special choice of the
parameters one gets solitons that, as a result of their mutual interaction,
exhibit a two dimensional shift and also a change of form.
If we require, in addition, to the representative point $p$ of the
two-soliton solution to belong to the hyperplanes of lower dimensions
$\lambda_{n}=\lambda_{m}$ or $\mu_{n}=\mu_{m}$ the two solitons,
because of their mutual interaction, not only are shifted in the plane
and change their form but also exchange mass. Surprisingly enough, in
both cases the relevance of the bifurcation effect depends on the
relative initial position of the two solitons.

In order to describe a generic soliton with parameters $\lambda$,
$\mu$ and $\gamma={1\over4}\eta\rho$ we introduce, in agreement with
the notation for the one soliton solution, the two variables
\beqs
\xi_1&=&-\mu_{\Im}(u-u_0)-\lambda_{\Im}(v-v_0)+2(\lambda_{\Im}\lambda_{\Re}+
\mu_{\Im}\mu_{\Re})t\nonumber\\
\xi_2&=&\mu_{\Im}(u-u_0)-\lambda_{\Im}(v-v_0)+2(\lambda_{\Im}\lambda_{\Re}-
\mu_{\Im}\mu_{\Re})t
\eeqs
the phase
\begin{equation}
\phi=\mu_{\Re}(u-u_0)+\lambda_{\Re}(v-v_0)+
(\lambda_{\Im}^2-\lambda_{\Re}^2+\mu_{\Im}^2-\mu_{\Re}^2)t
\end{equation}
and the functions
\beqs
&&a^{(\pm)}=\gamma\exp(\pm\xi_1),\qquad
b^{({+})}=(1+\gamma)\exp(\xi_2),\qquad
b^{({-})}=\gamma\exp(-\xi_2)\qquad\nonumber\\
&&D(\xi_1,\xi_2)=a^{({+})}+a^{({-})}+b^{({+})}+b^{({-})}.
\eeqs
To any of these variables and functions we add a label $(n)$ when we
are using the parameters $\lambda_n$, $\mu_n$ and
$\gamma_n={1\over4}\eta_n\rho_n$ of the $n$th soliton.

Then the two--soliton solution can be written as
\begin{equation}
q=-2i{N\over\Delta}
\end{equation}
where
\beqs
N&=&{1\over2}\eta_1{\lambda\cut}_{11}\exp(i\phi_{(1)})
\left[|{\lambda\cut}_{12}|^2|{\mu\cut}_{12}|^2b^{({+})}_{(2)}+
{\mu\cut}_{12}\mu_{12}{\lambda\cut}_{21}{\overline{\lambda}}_{21}
b^{({-})}_{(2)}+\right.\nonumber\\
&&\left.|{\lambda\cut}_{12}|^2{\mu\cut}_{12}
\mu_{12}a^{({+})}_{(2)}+ |{\mu\cut}_{12}|^2
{\lambda\cut}_{21}{\overline{\lambda}}_{21}
a^{({-})}_{(2)}\right]+(1\leftrightarrow 2)\\
\Delta&=&|\lambda_{12}|^2|\mu_{12}|^2b^{({-})}_{(1)}b^{({-})}_{(2)}
+|\lambda_{12}|^2|{\mu\cut}_{12} |^2\left[b^{({-})}_{(1)}a^{({-})}_{(2)}+
a^{({-})}_{(1)}b^{({-})}_{(2)}+a^{({-})}_{(1)}a^{({-})}_{(2)}\right]+
\nonumber\\
&&|{\lambda\cut}_{12}|^2|\mu_{12}|^2\left[b^{({-})}_{(1)}a^{({+})}_{(2)}+
a^{({+})}_{(1)}b^{({-})}_{(2)}+a^{({+})}_{(1)}a^{({+})}_{(2)}\right]+
\nonumber\\
&&|{\lambda\cut}_{12}|^2|{\mu\cut}_{12} |^2
\left[b^{({+})}_{(1)}b^{({+})}_{(2)}+
a^{({+})}_{(1)}b^{({+})}_{(2)}+b^{({+})}_{(1)}a^{({+})}_{(2)}
+a^{({-})}_{(1)}b^{({+})}_{(2)}+\right.\nonumber\\
&&\left.b^{({+})}_{(1)}a^{({-})}_{(2)}+b^{({-})}_{(1)}b^{({+})}_{(2)}+
b^{({+})}_{(1)}b^{({-})}_{(2)}+
a^{({+})}_{(1)}a^{({-})}_{(2)}+a^{({-})}_{(1)}a^{({+})}_{(2)}\right]-
\nonumber\\
&&{1\over2}\mbox{Re}\,\left[\eta_1\overline{\eta}_{2} {\lambda\cut}_{11}
{\lambda\cut}_{22}{\lambda\cut}_{21}{\mu\cut}_{12} \exp(i(\phi_{(1)}-
\phi_{(2)}))\right]
\eeqs
with
\begin{equation}
\mu_{12}=\mu_1-\mu_2, \qquad \lambda_{12}=\lambda_1-\lambda_2.
\end{equation}
We consider the discrete values $\lambda_n$ fixed, with for instance
$\lambda_{2\Re}>\lambda_{1\Re}$, and we study  the bifurcations of the
solution at large time with respect to the parameters $\mu_n$. There is
a bifurcation at $\mu_{1\Re}=\mu_{2\Re}$ and at $\mu_{1}=\mu_{2}$.
We need therefore to compute separately the asymptotic behaviour of the
two--soliton solution in the two cases. In particular we study the
asymptotic behaviour $q^{(\mp)}_{(2)}(u,v)$ of the second soliton at
$t=\mp\infty$. Thanks to the symmetry of the two--soliton solution the
asymptotic behaviour of the other soliton is simply obtained by
exchanging in the formulae the labels 1 and 2.

For $\mu_{1\Re}=\mu_{2\Re}$ we get
\beqs
&&q^{({-})}_{(2)}=-2\eta_2\lambda_{2\Im}
\exp[i(\phi_{(2)}{-}\phi_{(2)}^{({-})})]\frac{1{+}
\exp(-i\phi_0^{(2)})E_{(2)}}{D_{(2)}{+}E_{(2)}{\widetilde{D}}_{(2)}}
\left(\xi_1^{(2)}{-}\xi_0^{(2)({-})},\xi_2^{(2)}{-}\xi_0^{(2)({-})}\right)
\nonumber\\
&&q^{({+})}_{(2)}=-2\eta_2\lambda_{2\Im}
\exp[i\phi_{(2)})]\frac{1+\frac{\gamma_1}{1+\gamma_1}
\exp(-i\phi_0^{(2)})E_{(2)}}{D_{(2)}{+}
\frac{\gamma_1}{1+\gamma_1}E_{(2)}{\widetilde{D}}_{(2)}}
\left(\xi_1^{(2)},\xi_2^{(2)}\right)
\eeqs
with
\beqs
&&E_{(2)}\left(\xi_1^{(2)},\xi_2^{(2)}\right)=
\exp\left[\frac{\mu_{1\Im}}{\mu_{2\Im}}\left(\xi_1^{(2)}-\xi_2^{(2)}\right)
+\xi_0^{(2)}\right]c_{12}\nonumber\\
&&c_{12}=\exp[2\mu_{1\Im}(u_{01}-u_{02})]\nonumber\\
&&{\widetilde{D}}_{(2)}\left(\xi_1^{(2)},\xi_2^{(2)}\right)=
D_{(2)}\left(\xi_1^{(2)}+\xi_0^{(2)},\xi_2^{(2)}-\xi_0^{(2)}\right)\nonumber\\
&&\xi_0^{(2)}=\ln\left|\frac{\mu_{12}}{{\mu\cut}_{12}}\right|,\qquad
\phi_0^{(2)}=\arg\overline{\mu}_{12}{\mu\cut}_{12}\nonumber\\
&&\xi_0^{(2)({-})}=
\ln\left|\frac{\lambda_{12}}{{\lambda\cut}_{12}}\right|,\qquad
\phi^{({-})}_{(2)}=\arg\overline{\lambda}_{12}{\lambda\cut}_{12}.
\eeqs
The mass of the second soliton at $t=\pm\infty$ is
\begin{equation}
M^{(\pm)}_{(2)}=4\ln\frac{1+\gamma_2}{\gamma_2}.
\end{equation}
Therefore the position and the phase of the soliton are shifted, its
shape is changed but it does not exchange mass with the other
soliton. Note that its shape depends on the relative initial position
of the two solitons.

For $\mu_1=\mu_2$ we get
\beqs
&&q^{({-})}_{(2)}=\frac{-2\eta_2\lambda_{2\Im}
\exp[i(\phi_{(2)}{-}\phi_{(2)}^{({-})})]}
{\left[\left(1{+}c_{12}\frac{1+\gamma_2}{\gamma_2}\right)
a_{(2)}^{({+})}
{+}a_{(2)}^{({-})}{+}b_{(2)}^{({+})}{+}(1{+}c_{12})b_{(2)}^{({-})}\right]
\left(\xi_1^{(2)}{-}\xi_0^{(2)({-})},\xi_2^{(2)}{-}\xi_0^{(2)({-})}\right)}
\nonumber\\
&&q^{({+})}_{(2)}=\frac{-2\eta_2\lambda_{2\Im}
\exp[i\phi_{(2)}]}
{\left[\left(1{+}c_{12}\frac{\gamma_1}{\gamma_2}
\frac{1+\gamma_2}{1+\gamma_1}\right) a_{(2)}^{({+})}
{+}a_{(2)}^{({-})}{+}b_{(2)}^{({+})}{+}\left(1{+}c_{12}
\frac{1+\gamma_1}{\gamma_1}\right)b_{(2)}^{({-})}\right]
\left(\xi_1^{(2)},\xi_2^{(2)}\right)}
\eeqs
and
\beqs
&&M^{({-})}_{(2)}=4\ln\frac{(1+\gamma_2)(1+c_{21})}
{c_{12}+\gamma_2(1+c_{21})}\nonumber\\
&&M^{({+})}_{(2)}=4\ln\frac{(1+\gamma_2)(1+\gamma_1+c_{21}\gamma_1)}
{\gamma_2(1+\gamma_1)+c_{21}\gamma_1(1+\gamma_2)}.
\eeqs
Therefore the soliton during the interaction shifts its position,
changes shape and exchanges mass with the other soliton, while the total
mass of the two solitons is conserved. It is worth noting that the shape
and the energy of each soliton depend on the relative initial position
of the two solitons.

Drawings describing the peculiar behaviours of solitons described in this
section can be found in \cite{bifurcationDubna90}.

Finally let us remark that also the boundary $a^{(2)}(u,t)$ bifurcates.
In fact, while in the generic case $a^{(2)}(u,t)$ describes in the
$(u,t)$ plane two infinite waves crossing at one point, for
$\mu_{1\Re}=\mu_{2\Re}$  it describes two parallel infinite waves and
for $\mu_1=\mu_2$ only one infinite wave.

\end{document}